\documentclass{emulateapj}
\usepackage{times}
\usepackage{apjfonts}
\usepackage{amssymb}
\usepackage{epsfig}
\usepackage{rotating}
\usepackage{amsthm}
\usepackage{rotating}
\usepackage{amsmath}

\def\hi{\textsc{Hi} }

 \def\us{\mathbf{u}}
\def\u{\mathbf{u},\nu}
\def\ut{\mathbf{u},\eta}
\def\sc{\vec{\theta}, \nu}
\def\k{\mathbf{k}}

\def\kperp{k_{\bot}}
\def\kpar{k_{\|}}

\begin{document}

\title{Bright Source Subtraction Requirements For Redshifted 21~cm Measurements} 

\author{A. Datta\altaffilmark{1,2}, J.D. Bowman \altaffilmark{3,4} and
C.L. Carilli \altaffilmark{2}, } 

\altaffiltext{1}{New Mexico Tech, Socorro, NM 87801, USA}
\altaffiltext{2}{National Radio Astronomy Observatory, Socorro, NM
  87801, USA}
\altaffiltext{3}{California Institute of Technology, Pasadena, CA
  91125, USA} 
\altaffiltext{4}{Hubble Fellow}

\email{adatta@nrao.edu}

\begin{abstract}
The \hi 21~cm transition line is expected to be an important probe into
the cosmic dark ages and epoch of reionization. Foreground source
removal is one of the principal challenges for the detection of this
signal. This paper investigates the extragalactic point source
contamination and how accurately bright sources ($\gtrsim 1$ ~Jy) must
be removed in order to detect 21~cm emission with upcoming radio
telescopes such as the Murchison Widefield Array (MWA). We consider
the residual contamination in 21~cm maps and power spectra due to
position errors in the sky-model for bright sources, as well
as frequency independent calibration errors. We find that a source
position accuracy of $0.1$~arcsec will suffice for detection of the
\hi power spectrum. For calibration errors, $ 0.05~\%$ accuracy
in antenna gain amplitude is required in order to detect the cosmic
signal. Both sources of subtraction error produce residuals that are 
localized to small angular scales, $\kperp \gtrsim 0.05 $~Mpc$^{-1}$,
in the two-dimensional power spectrum.

\end{abstract}

\keywords{Cosmology: Early Universe, Galaxies: Intergalactic Medium,
Radio Lines: General, Techniques: Interferometric, Methods: Data
Analysis}

\section{Introduction}
The cosmological epoch of reionization (EoR) is a key milestone in the
history of structure formation, marking the transition from a fully
neutral to a highly ionized intergalactic medium (IGM) due to the
ultra-violet and X-ray radiation of early stars, galaxies, and black
holes.  Recent observations of the Gunn-Peterson effect, i.e.,
Ly$\alpha$ absorption by the neutral IGM, toward the most distant
quasars ($z \sim 6$), and the large scale polarization of the CMB,
corresponding to Thompson scattering during reionization, have set the
first constraints on the reionization process. These results suggest
significant variance in both space and time, starting perhaps as far
back as $z \sim 11$ (\citet{komatsu10}; WMAP 7 year data) and
extending to $z \sim 6$ \citep{fan06b}. Previous WMAP five year data
indicates the 5~$\sigma$ detection of the E-mode of polarization which
rules out any instantaneous reionization at $z \sim 6$ at 3.5~$\sigma$
level. In case of the Gunn-Peterson effect, the IGM becomes optically
thick to Ly$\alpha$ absorption for a neutral fraction as small as
$\sim 10^{-3}$. In order to overcome these limitations, it has been
widely recognized that mapping the red-shifted \hi 21~cm line has great
potential for direct studies of the neutral IGM during reionization 
\citep{barkana01, fan06b, furlanetto06, morales09}. 

There are number of upcoming low-frequency arrays with key science
goals to detect the \hi 21~cm signal from the EoR. This includes the
Murchison Widefield Array [MWA] \citep{mitchell08, lonsdale09, bowman06},
Precision Array to Probe Epoch of Reionization [PAPER]
\citep{backer07, parsons09a}, Low Frequency Array [LOFAR]
\citep{harker10, jelic08, panos09} and Giant Meterwave Radio Telescope
[GMRT] \citep{pen09}.  One of the major challenges for all of these
upcoming arrays will be the removal of the continuum foreground
sources in order to detect the faint \hi signal from the EoR.  

A variety of continuum foregrounds complicate redshifted 21~cm
measurements of the EoR \citep{shaver99}. Diffuse Galactic synchrotron
emission dominates the low-frequency radio sky and is approximately
four orders of magnitude brighter than the $\sim10$~mK 21~cm signal at
the frequency relevant to reionization ($\nu\approx150$~MHz).  In
addition, Galactic and extragalactic free-free emission contribute
additional flux to the diffuse foreground.  Radio point sources from
AGN, radio galaxies, and local Galactic sources are numerous and
particularly challenging.  The brightest of these sources have fluxes
well above $S>1$~Jy and are seven or eight orders of magnitude above
the EoR signal in low-frequency radio maps.  The distribution of point
sources also extends to very faint levels such that the brightness
temperature due to confused sources in upcoming arrays will be
$\sim10$~K, or three orders of magnitude brighter the than the 21~cm
background.   

In this paper we discuss how the radio interferometric imaging
techniques are going to affect the foreground source modeling and
subsequent removal from the data-set in order to search for the EoR
signal. Recently, there has been extensive research on foreground
source modeling at these low frequencies \citep{dimatteo02, jelic08,
  rajat09}. Similar effort has also been made in exploring different
techniques to remove the foregrounds from the EoR data-set by
\citet{morales06a,morales06b, bowman09, liu09, parsons09b, gleser08,
  harker09a, harker09b}. Since attempting to observe a signal below
the confusion limit of foreground sources is a novel aspect of 21~cm
experiments, most of these works primarily focus on the removal of
faint and confused sources that fall below a specified cutoff flux
limit, $S_{cut}$ ($\approx 1$~Jy). They do not consider the foreground 
sources brighter than $S_{cut}$ and how accurately they need to be
removed.  Indeed, most of these analysis implicitly assume that the
bright foreground sources above $S_{cut}$ have been removed perfectly.
But in reality imperfect instrument calibration or any errors in the
subtracted foreground model will introduce artifacts and leave
residual contamination in the data after bright source removal, even
by traditional techniques such as ``peeling''.  These residuals may
interfere with either the subsequent faint source subtraction or the
ultimate detection and characterization of the redshifted 21~cm
signal.  

In \citet{abhi09} we dealt with the bright point sources above
$S_{cut}$ and the limitations that will be caused due to imperfect
removal of such sources in the image plane. In this paper we extend
the initial analysis in order to estimate the residual contamination
in the power spectral domain of improper bright source subtraction.
The objective of this paper is to demonstrate how the accuracy in the
foreground removal affects the detection of \hi 21~cm power spectra
with the MWA.  In Section 2, we discuss our choice of sky model and
outline the simulation parameters, including the array specifications
and data reduction procedure, and describe the two categories of
corruption terms that we will consider: source position errors and
residual calibration errors. The results obtained for the residual
angular power spectrum, spherically averaged three-dimensional power
spectrum, and two-dimensional power spectrum are presented in
Section~3. Finally, in the last section we summarize the implications 
of the results from our simulations.

\section{THE SIMULATION}

\subsection{Sky Model}

Our main aim is to explore the level of accuracy needed in instrument
calibration and foreground modeling in order to ensure the residual
errors from bright foreground source removal do not obscure the
detection of the signal from cosmic reionization. With this goal in
mind, we use a simple sky model for our simulations that only includes
bright radio point sources.  No diffuse emission from the Galaxy is
included as a part of the sky model; and the 21~cm signal and thermal
noise are also omitted.  Our sky model is derived from the $\log N$ --
$\log S$ distribution of sources and is termed the ``Global Sky
Model'' (GSM) from now onwards. Since the GSM only includes sources
above 1~Jy, we follow the source counts from the 6C survey at 151~MHz
\citep{hales88}:   

\begin{equation}
N(>S_{Jy}) = 3600~ S_{Jy}^{-2.5} Jy^{-1} str^{-1}, \label{e_gsm}
\end{equation}
For a field-of-view of $15^\circ$ the total number of sources above 1~Jy
is $\sim 170$, following the above power-law distribution. The
entire flux range, between 1~Jy and $10^3$~Jy, has been subdivided
into several bins (in logarithmic scale) and populated with the number
of sources that corresponds to the flux range of each bin (according to
Equation~ \ref{e_gsm}). Inside each bin, we have assigned each source
a flux density following a normal distribution. The strongest source
in our GSM is $\sim 200$~Jy.  The observed distribution of radio
sources shows evidence for only very weak angular clustering and the
brightest extragalactic sources in the sky are not clustered at all
\citep{blake02}. Therefore, in order to assign a position to each of 
these sources within the field-of-view, a uniform random number
generator has been used which predicted the offset from the field
center for respective sources. In the GSM all the foreground sources
are flat spectrum, i.e. with zero spectral index ($\alpha = 0$). 

\begin{figure}[h!]
\epsscale{1}
\plotone{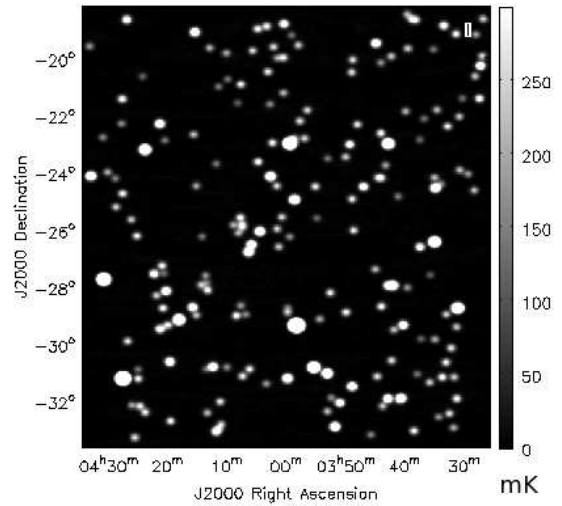}
\caption{Simulation of the sky model centered on RA=4h and
  DEC=$-26^\circ$ as would be observed by the MWA. Clark-CLEAN has
  been applied to this image using w-projection (256 planes) and
  natural weighting \citep{abhi09}.}  
\label{f_model} 
\end{figure}

Figure~\ref{f_model} shows a simulated image of our bright source sky
model that has been produced using the procedure described below.  A
wide-field variant of the well-known Clark-CLEAN algorithm that
utilizes a w-projection algorithm for 3-dimensional imaging was
applied to the simulated image.  The apparent angular size of the
sources in Figure~\ref{f_model} reflects the size of the synthesized
beam. The input sources in the model are treated as ideal point
sources. This input sky model is used for all the simulations
presented in this paper.  

\subsection{Array Specifications}
\label{s_spec}

\begin{figure}[h!]
\epsscale{1}
\plotone{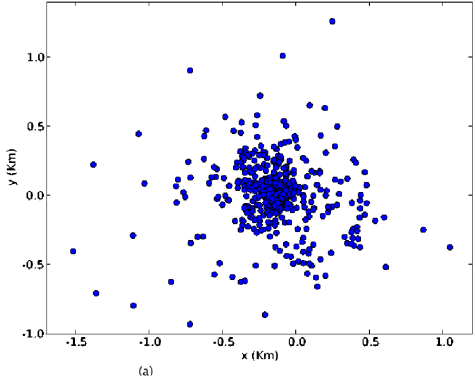}
\caption{Array layout for the 512~elements with maximum baseline
of 1.5~km. }
\label{f_layout}
\end{figure}

Table~1 outlines the instrumental parameters that we have assumed for
this analysis. Most of these parameters reflect  the current
specifications for the MWA, but we note that the array is presently
under development and some properties may be subject to change.   In
addition, we have intentionally reduced the simulated field of view
compared to the actual MWA in order to reduce the computational
overhead of the simulation.  Figure~\ref{f_layout} shows the array
layout for the 512 element array with maximum baseline of 1.5~km.  

\begin{deluxetable}{cc}
\tablecaption{Array Specifications \label{tab_detector} }
\tablehead{\colhead{Parameters} &  \colhead{Values}}
\tablecolumns{2}
\startdata
No. of Tiles                    & 512 \\ \hline
Central Frequency               & 158 MHz ($z \sim 8$) \\ \hline
Field of View                   & $\sim$ 15$^o$ at 158 MHz. ($\propto \lambda$)  \\ \hline
Synthesized beam                & $\sim$ 4.5' at 158 MHz. ($\propto \lambda$) \\ \hline
Effective Area per Tile         & $\sim$ 17 $m^2$ \\ \hline
Maximum Baseline                & $\sim$ 1.5 km \\ \hline
Total Bandwidth                 & 32 MHz \\ \hline
$T_{sys}$                       & $\sim$ 250 K \\ \hline
Channel Width                   & $\sim$ 32 kHz \\ \hline
Thermal Noise                   & $\sim$ 7.55 mK \\
      ~ ~                       & (5000 hours $\&$ $2.5$ MHz) \\
\enddata
\tablecomments{Array parameters have been influenced by the MWA
specifications as mentioned in \citet{mitchell08} and
\citet{bowman09}. Original MWA Field-of-view is $\sim 25^{o}$ at 150 MHz.} 
\end{deluxetable}

For the purposes of modeling earth rotation synthesis in the
instrumental response, the center of the target field is chosen such
that it coincides with one of the cold spots in the foreground
Galactic synchrotron emission visible from the southern hemisphere
location of the MWA.  The exact field center used for the GSM is 4
hours in Right Ascension and -26 degree in Declination.  Most of the
upcoming low-frequency telescopes, including the MWA, will only be
able to observe a field around its transit.  We have used 6 hours of
integrations for all the simulations, assuming that the telescopes
will observe a field between $\pm$~ 3 hours in Hour Angle from
transit. 

\begin{figure*}[t!]
\centering
\epsfig{file=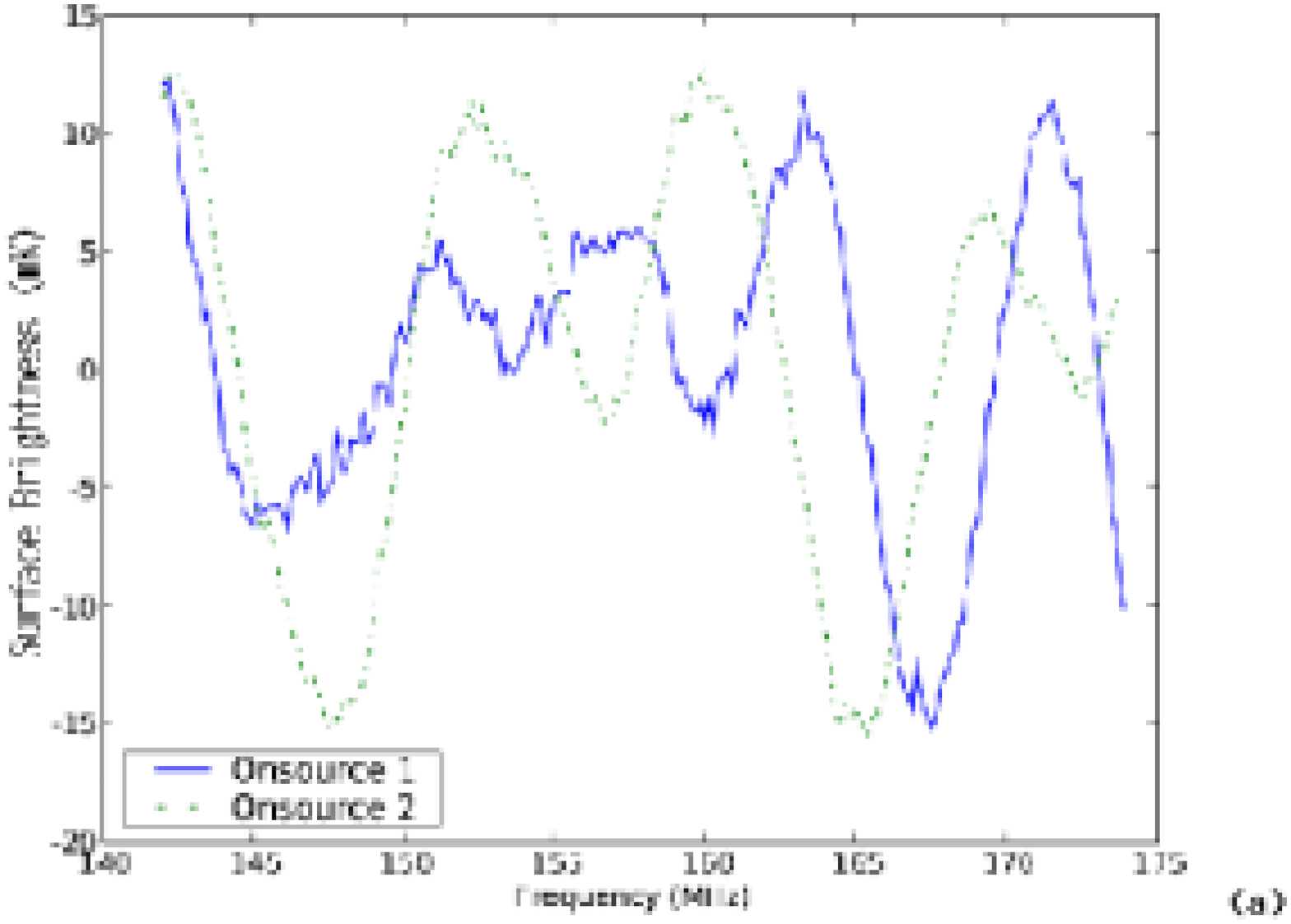,height=2.6truein}
\epsfig{file=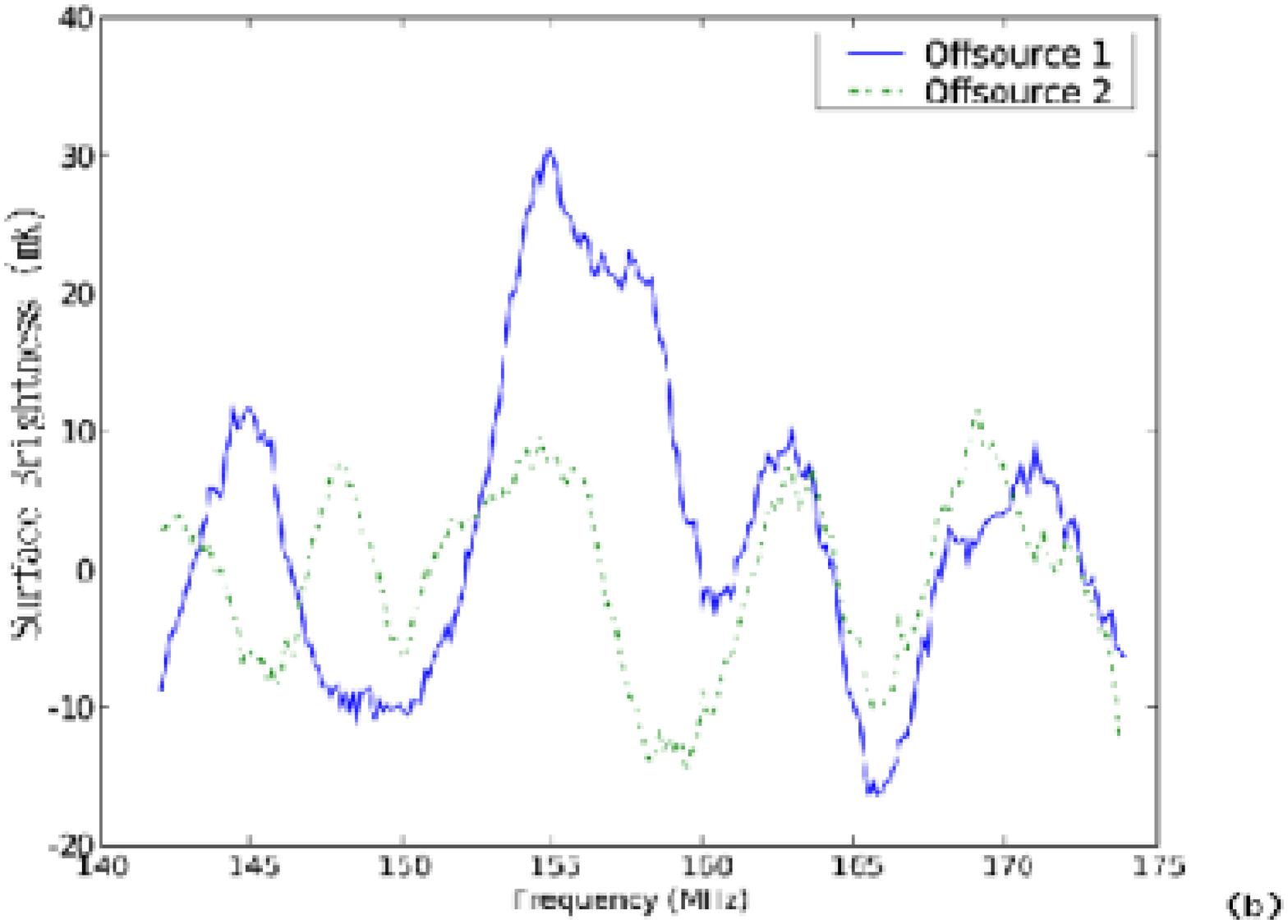,height=2.6truein}
\caption{({\bf a}) The spectral profile along two lines of sight in
  the the final residual image cube following the IMLIN step.  In
  this case, the two pixels were chosen to be next to the positions of
  sources in the input model. ({\bf b}) Same as panel (a), but here,
  the two pixels were chosen to be far from any sources in the input
  model. Synthesized beam area of 4.5 arcmin $\times$ 4.5 arcmin is
  used to convert the flux densities to surface brightness.}   
\label{f_profile} 
\end{figure*}
\subsection{Data Reduction Procedure}
\label{sec_datared}

The $15^o$ field-of-view will include $\sim 170$ bright sources ($ >
1$~Jy). The individual flux densities of these foregrounds are $\sim
10^5 - 10^7$ times higher than the signal from cosmic reionization
that these instruments are aiming to detect. So the challenge lies in
calibration and subsequent removal of such bright sources from the raw
data-sets. The data rate of 19 GB/s \citep{mitchell08} will not allow
the MWA to store the raw visibilities produced by the
correlator. Hence real-time calibration and imaging needs to be done
in order to reduce the data volume and store the final product in the
form of image cubes \citep{mitchell08}. The critical steps include
removal of the bright sources above the $S_{cut}$ level from the
data-sets in these iterative rounds of real-time calibration and
imaging procedure. As a result the residual image-cubes will not be
dominated by these bright sources and the rest of the foregrounds can
be removed in the image domain.      

However, the accuracy of the foreground source removal strategies are
strongly dependent on the data reduction procedure. The likely data
reduction procedure which will be followed by the upcoming telescopes
can be broadly outlined as :

\begin{itemize}
\item The raw data-sets from the correlator will go through real-time
calibration and subsequent removal of the bright sources based on some
Global Sky Model (GSM), down to $S_{cut}$ level, in the UV domain. 
\item The residual data-sets will be imaged and stored as a cube for
the future processing and removal of sources which are below
$S_{cut}$. 
\end{itemize}

\begin{figure}[h!]
\epsscale{1}
\plotone{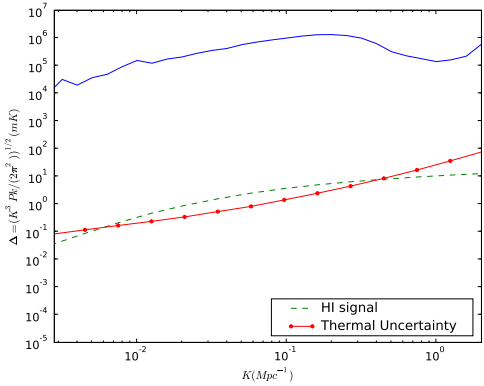}
\caption{1D spherically averaged power spectrum of the input Global
  Sky Model showing the total power of the bright sources in the sky
  model before any foreground removal has been applied. The thermal
  noise uncertainty for a 300 hour observation by the MWA is also
  shown, along with the \hi 21~cm signal power spectrum for a fully
  neutral IGM ($x_{\hi}=1$) at $z=8$ \citep{furlanetto06}.} 
\label{f_pow_tot}   
\end{figure}

The simulated data reduction pathway that we follow in this paper is:

\textit{ {\bf (i)}} First, the observed visibilities
($V_{ij}^{Obs}(\u) \equiv V_{ij}^{GSM_{perfect}}(\u)$) are simulated
for a 6~hour observation ($\pm 3$~hours in Hour-angle) using the GSM
and the array configuration from Section~\ref{s_spec}. In the above
notation, $\u \equiv (u,v)$ denotes the Fourier conjugate of the sky
coordinate ($\theta_x,\theta_y$) and $\nu$ is the frequency of
observations.       

\textit{ {\bf (ii)}} Next, we generate the foreground model
($V_{ij}^{mod}(\u)$) that will be subtracted from the observation.  In
this case, the model is corrupted to either simulate errors in the
assumed positions of the sources or to simulate calibration errors.
For the source position errors, the model visibilities are given
simply by: 
\begin{equation}
V_{ij}^{mod}(\u) = V_{ij}^{GSM_{imperfect}}(\u),
\end{equation}
where the position of each source has been slightly moved from its
original location by a distance drawn from a Gaussian distribution
with standard deviation $\sigma_{\theta}$.  We assume that the source
position errors are constant throughout the entire duration of the
observation, as would be the case for a foreground model constructed
from either an outside catalog or from the data itself at the
conclusion of the observation.  This is an idealization that may be
broken in practice if sources are ``peeled'' in real-time.  

For the residual calibration errors, the model visibilities are given by:
\begin{equation}
V_{ij}^{mod}(\u) = g_i(t)g_j^{*}(t)V_{ij}^{GSM_{perfect}}(\u), \label{e_cal}
\end{equation}
where $g_i(t) \approx (1+a_i)e^{i\phi_i}$ are the antenna-dependent
complex gains.  The parameters $a_i$ and $\phi_i$ denote small
amplitude and phase deviations, respectively, and are each drawn from
there own Gaussian distribution with standard deviation $\sigma_a$ or
$\sigma_{\phi}$ \citep{abhi09}.  

The MWA will produce calibration solutions in real-time with an
$\sim8$~second cadence.  This rapid pace is planned in order to
simultaneously calibrate both the instrument hardware properties and
the ionospheric phase screen.  It is not known, yet, if in practice
the residual calibration errors from the real-time processing will be
largely independent or highly correlated between individual 8-second
solutions.  This is an important experimental property to consider in
our simulation, because the degree of correlation greatly affects the
level of accuracy needed in individual calibration solutions.  If the
individual errors are largely independent, then each 8-second sample
can be modeled as coming from a Gaussian distribution and the accuracy
tolerance will be relatively loose since many samples will be
available and tend to average toward zero.  Such a situation would be
the best-case scenario.  On the other hand, if the calibration errors
are highly correlated, then each calibration solution must meet a much
more stringent accuracy level to achieve the same residual
contamination at the end of the full integration.    

For our simulation procedure, we assume a relatively conservative
scenario that the residual errors in a given antenna's 8-second
calibration solutions are perfectly correlated for the duration of one
6-hour observing night, but perfectly uncorrelated between successive
observing nights.  We further assume that the residual errors {\it
  between} antennas are perfectly uncorrelated at all times.  This
choice is somewhat arbitrary given the current level of knowledge, but
we believe it is a plausible fiducial case since both the overall
ionospheric properties and the ambient conditions may change
significantly from day-to-day.  Hence, in our simulation, $\sigma_a$
and $\sigma_{\phi}$ are used to draw a calibration error value ($a_i$
and $\phi_i$) from a Gaussian distribution only once per antenna per
night and that specific error is applied to all the simulated 8-second
solutions for the given antenna throughout the 6-hour period of
rotation synthesis. When the next night's observing block commences, a
new error is drawn from the distribution for each antenna, and so
on. 

\textit{ {\bf (iii)}} Now we are ready to calculate the residual
visibilities by subtracting the foreground model produced in step (ii)
from the simulated observation of step (i) according to: 
\begin{equation}
V_{ij}^{res}(\u)=V_{ij}^{obs}(\u)-V_{ij}^{mod}(\u).  \label{e_res}
\end{equation}
In the sections below, we refer to this step as ``UVSUB'' since it
was implemented using the UVSUB algorithm \citep{cornwell92}. For the residual
calibration errors, we can reduce Equation~\ref{e_res} by substituting
in with Equation~\ref{e_cal} and simplifying to obtain: 
\begin{equation}
V_{ij}^{res}(\u) = \left(1 - g_i g_j^*\right) V_{ij}^{GSM_{perfect}}
\end{equation}

\textit{ {\bf (iv)}} At this point, we have completed the subtraction
of bright sources from our simulated observation, leaving only
residual contamination due to the differences between our simulated
observation and the corrupted model.  Example images of the residual
contamination at this stage are shown in \citet[panels (a) of their
  Figures 5, 7, and~9]{abhi09}. In practice, this bright
source-subtracted data cube will be the starting point for the second 
stage of redshifted 21~cm foreground subtraction that aims to remove
faint and confused sources by fitting and subtracting a low-order
polynomial along the frequency axis for each line of sight in the data
cube.  We want to understand how this additional process affects the
end result of the bright source removal, so we approximate the faint
source polynomial fitting here by applying a Fourier transform to the
UV map generated from the residual visibilities in order to produce a
residual dirty image cube, $I^{res}$.  In this dirty image cube, we
fit a third order polynomial in frequency along each line of sight and
subtract it.  Thus, we obtain the final residual image,
$I^{res}_{IMLIN}(\sc)$.  We refer to this step as ``IMLIN'' for the
remainder of this paper because it was implemented with the IMLIN
algorithm \citep{cornwell92}. To illustrate this final result,
Figure~\ref{f_profile} shows residual spectral profiles along four
lines of sight in the dirty image cube after polynomial fitting and
subtraction.  Example images of the final residual contamination after
IMLIN are also shown in \citet[panels (b) of their Figures 5, 7,
  and~9]{abhi09}. 

Using higher order polynomials in the IMLIN step removes structures
at increasingly smaller scales \citep{mcquinn06}.  This improves the
foreground cleaning, but since the 21~cm reionization signal has
significant structures on scales that correspond to $\sim2.5$~MHz, or
approximately 10\% of the bandwidth over which the polynomial is fit,
it also has the potential to remove much of the 21~cm signal.  We have
restricted our attention to a third order polynomial in this
work because it is the lowest-order polynomial likely to be sufficient
for removing the faint continuum sources given their power-law
spectral shapes. 

We also explored using the UVLIN algorithm \citep{cornwell92}, which
fits and subtracts polynomials in the UV domain instead of the image
domain, eliminating the need to convert our residual data sets into
image cubes.  However, UVLIN works perfectly only within a small
field-of-view, depending on the channel width in frequency
\citep{cornwell92}, and was found to be inadequate for our purposes. 

\textit{ {\bf (v)}} The final step in our procedure is to calculate
power spectra from the residual image cubes and compare these residual
foreground power spectra to the theoretically predicted 21~cm power
spectrum and expected thermal noise power spectrum for the MWA.  We
calculate three forms of the residual power spectra from our final
data cubes: the derived angular power spectrum $C_{\ell}$ for a narrow
frequency channel, the spherically averaged three-dimensional power
spectrum $P(k)$ from the entire data cube, and the two-dimensional
power spectrum $P(\kperp,\kpar)$ found by averaging over transverse
modes in the full three-dimensional power spectrum.  Each of these
cases is discussed in more detail in Section~\ref{s_results}.  As a
reference, we show in Figure~\ref{f_pow_tot} the spherically averaged
power spectrum for our input GSM before any source removal has been
applied.    

In order to simulate the observed visibilities ($V_{ij}^{Obs}$), we
have used the simulator tool in the CASA software
\footnote{http://casa.nrao.edu/}. We have also used CASA to perform
the imaging and the subsequent IMLIN step. The rest of the operations
are performed using separately written python
\footnote{http://www.python.org/} scripts.

\section{RESULTS}
\label{s_results}

We begin our discussion of the results of the residual power spectrum
determination by reviewing our initial findings from \citet{abhi09}.
In that work, we explored the source position and calibration accuracy
needed to allow direct imaging of Stromgren spheres with very deep
integrations by the MWA.  Our simulations demonstrated that knowledge
of the true positions of the bright foreground sources in an MWA
target field is required to within $\sigma_{\theta}=0.1$~arcsec,
assuming Gaussian errors, in order for the residual contamination
following subtraction to be below the 21~cm signal from Stromgren
spheres in image maps that could be acquired by the MWA with
5000~hours of integration.  Similarly, in \citet{abhi09} we found
that, for the case of calibration errors corrupting the measurements
under the same conservative assumptions outlined in step (ii), a
calibration accuracy of $\sigma_{a}=0.2 \%$ in gain amplitude (or 
$\sigma_{\phi}=0.2$~degree in phase) is needed for the residual
contamination to be below the thermal noise in a part of the image map
far from any bright sources for a long integration by the MWA. 
 
Here, we focus our attention on the residual contamination that can be
tolerated in measurements of the power spectrum of a target field
observed by the MWA.  One of the primary motivations for seeking to
first detect and characterize the 21~cm power spectrum, rather than
immediately attempt to image the background, is that the MWA will only
require $\sim300$~hours of observing to have sufficient sensitivity to
detect the spherically-averaged 21~cm power spectrum at $z\approx8$,
assuming the IGM is not fully ionized at that time.  Because the power
spectrum measurement differs significantly from a direct imaging
observation, there are several key questions that we seek to address:
1) are the tolerances on the source position and calibration errors
greater (or lesser) than in the direct imaging case, 2) is a
particular region of the power spectrum likely to be more affected
by residual contamination than another, and 3) can we hope to build a
library of template models for foreground contamination that could be
used to marginalize out some of the contamination during the analysis
of the power spectrum? 

We will address these questions for each of the three classes of power
spectra listed in step (v) (section~\ref{sec_datared}) that the MWA is
likely to study: angular, spherically averaged, and two-dimensional.
For each class of power spectrum, we will present residual power
spectra for both corruption models: source position errors, and
calibration errors. And for each corruption model, we will use three
fiducial levels of error in our investigation: for source position
errors, our fiducial cases are $\sigma_{\theta}=\{$0.01, 0.1, and
1~arcsec$\}$, while for the calibration errors our fiducial levels are
$\sigma_a=\{$0.01, 0.1, and 1$\%\}$ in gain amplitudes, which also
translates to $\sigma_p=\{$0.01, 0.1, and 1$^\circ \}$ in phase
\citep{abhi09}. 

\subsection{Angular Power Spectrum}
\begin{figure*}[t!]
\centering
\epsfig{file=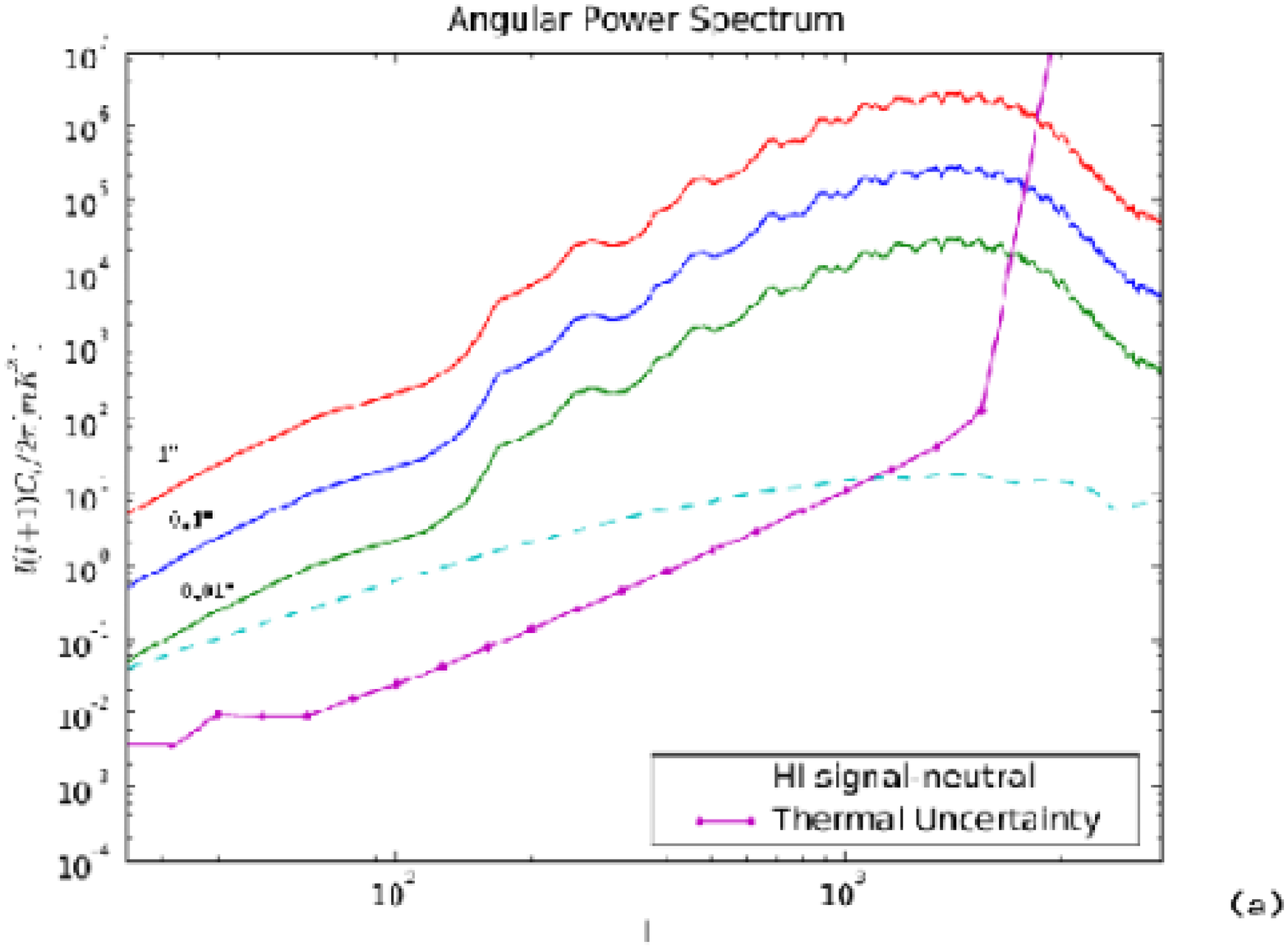,height=3.2truein}
\epsfig{file=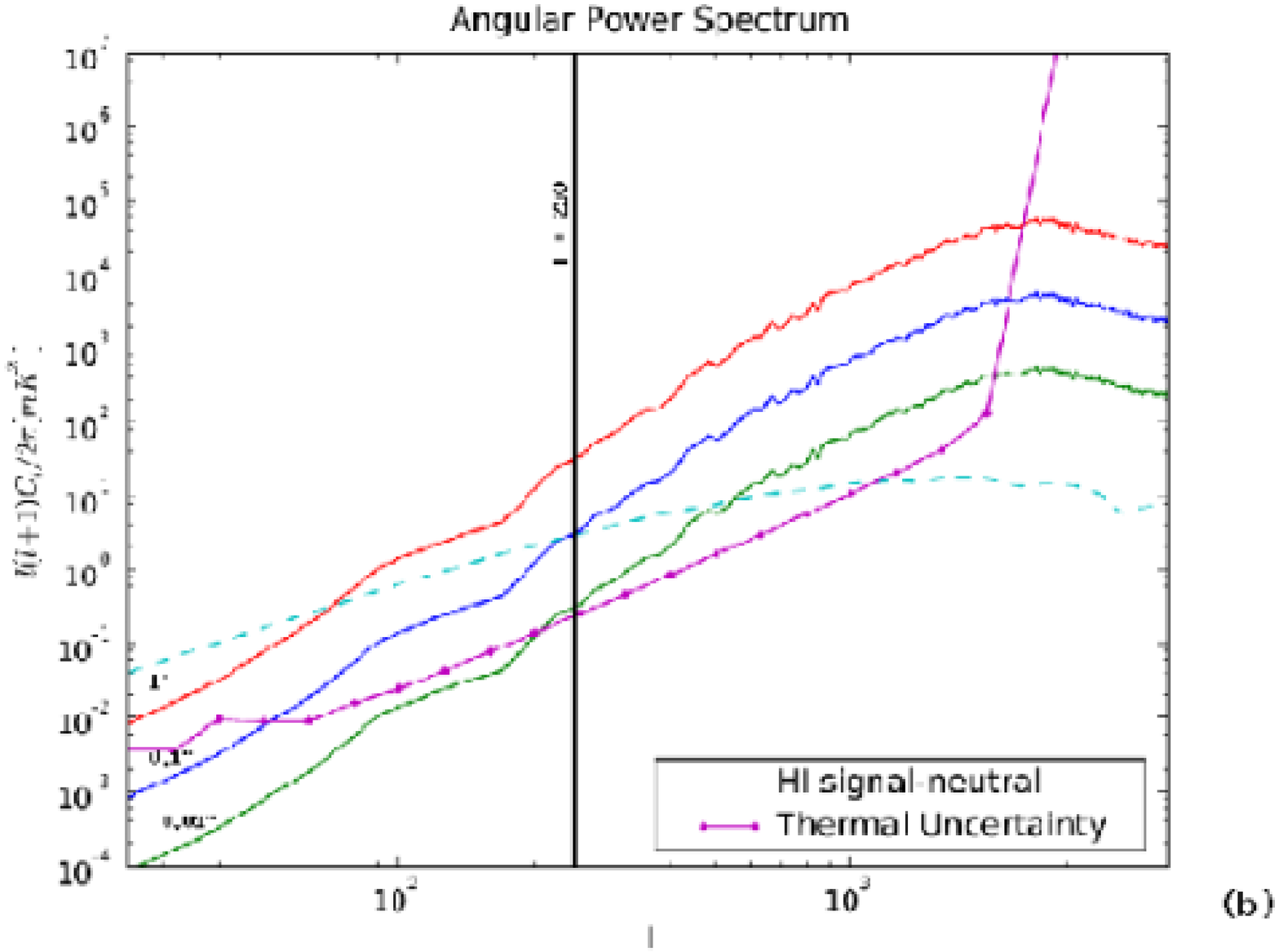,height=3.2truein}
\caption{({\bf a}) Angular power spectrum of the UVSUB residual
  image $I^{res}(\sc)$ made after subtraction of a foreground model
  with source position errors of $\sigma_{\theta}=\{$0.01, 0.1, and
  1~arcsec$\}$. ({\bf b}) Same as panel (a) but for the IMLIN
  residual image, $I^{res}_{IMLIN}(\sc)$, that is produced after
  polynomial fitting and subtraction has been applied along each
  sight-line. The vertical line in panel (b) corresponds to $\ell \sim
  250$, below which the polynomial fitting is expected to remove most
  of the structure.  Both panels include the thermal noise uncertainty
  power  spectrum assuming 5000 hours of observation with the MWA and the
  $\hi$ 21~cm signal power spectrum for a fully neutral IGM ($z\sim
  8$, $x_{\hi}=1$; \citet{furlanetto06}).  These angular power spectra
  are what would be expected from the MWA if it integrated deep enough
  to directly image a typical cosmic Stromgren Sphere.} 
\label{f_ang_src}
\end{figure*}

\begin{figure*}[t!]
\centering
\epsfig{file=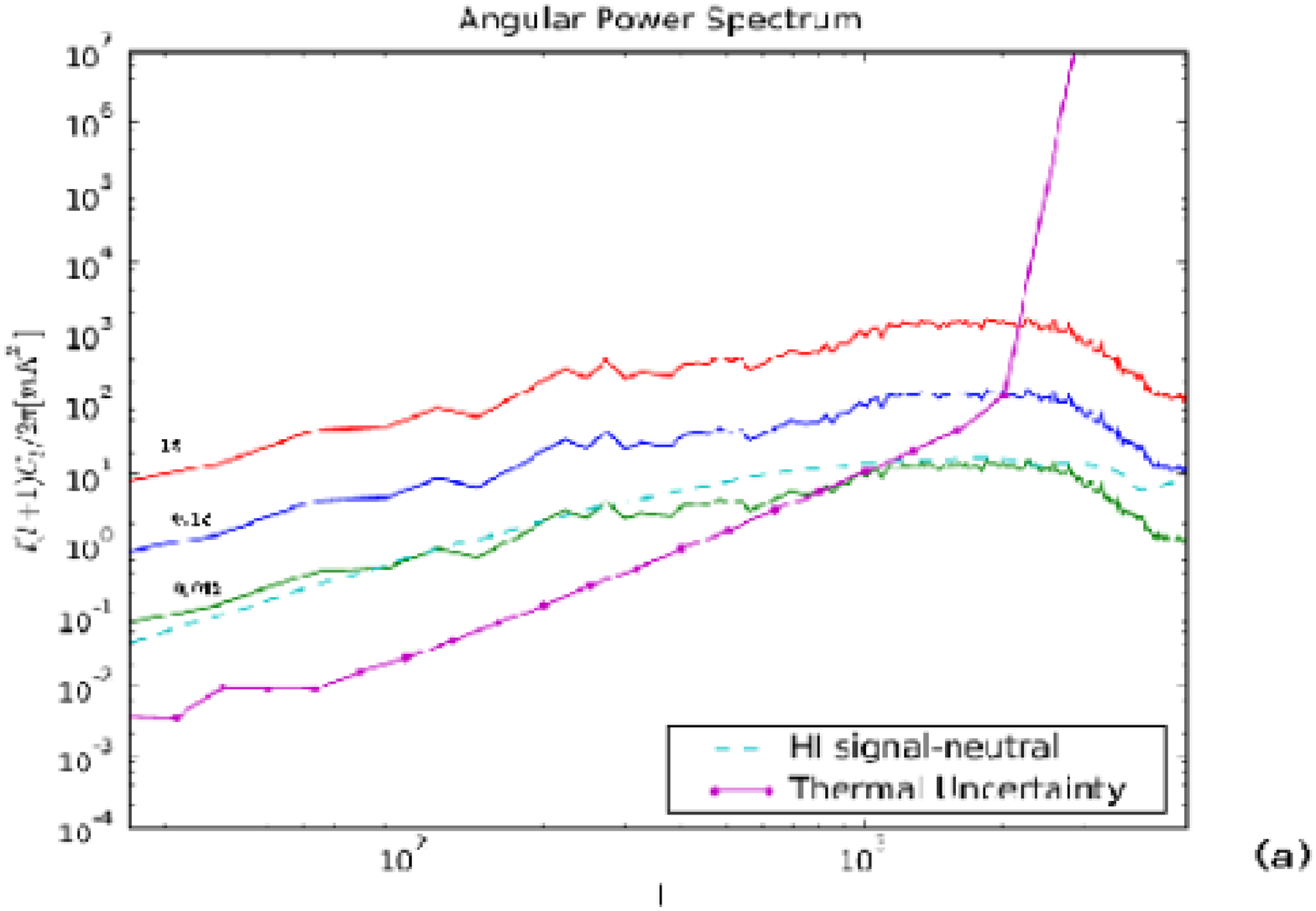,height=3.2truein}
\epsfig{file=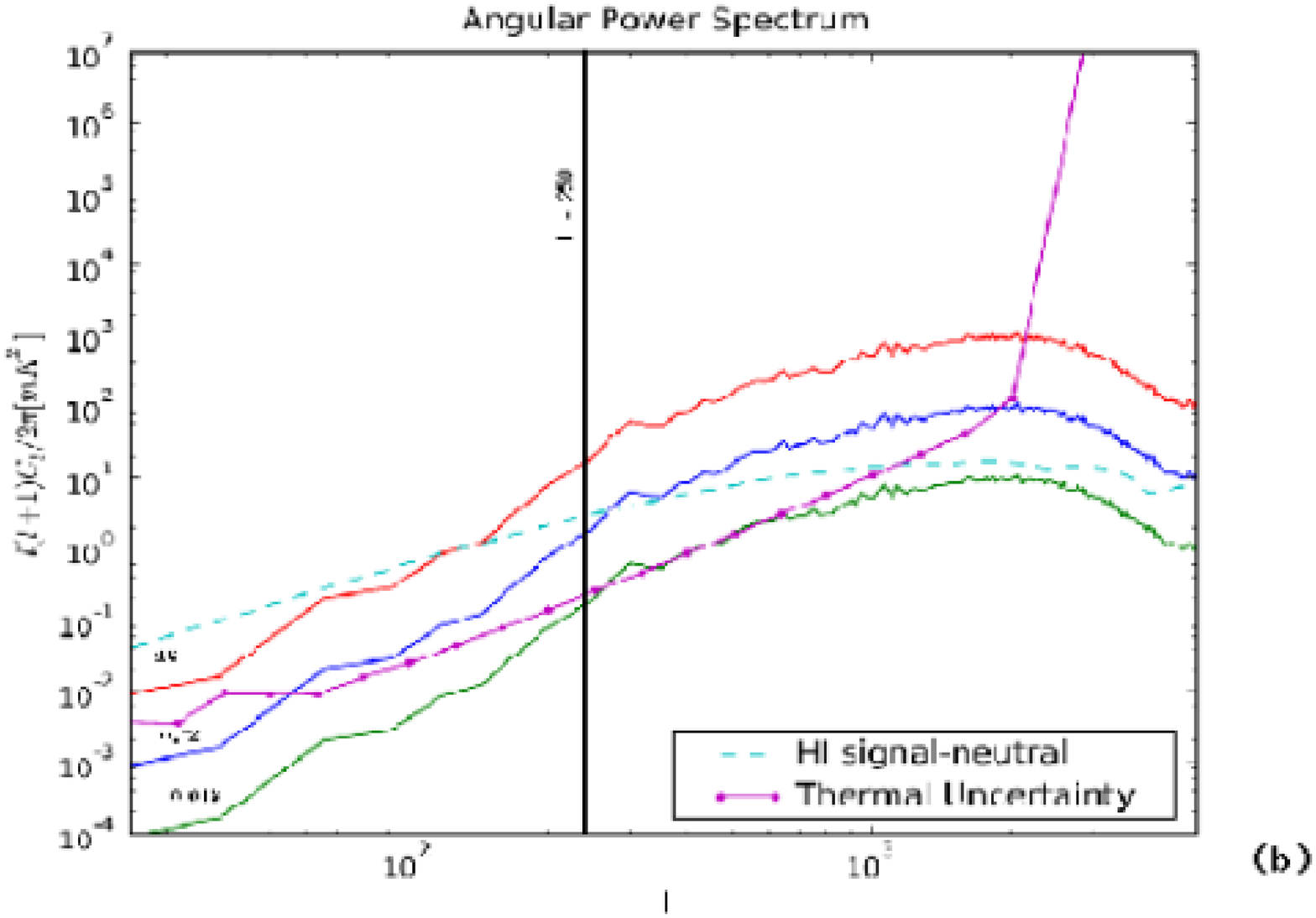,height=3.2truein}
\caption{Same as Figure~\ref{f_ang_src}, but for the residuals due to
  calibration errors.  In this case, the three residual angular power
  spectra are for errors of $\sigma_a=\{$0.01, 0.1, and 1$\%\}$.}  
\label{f_ang_cal}
\end{figure*}

\citet{white99} describe the technique to derive the angular power
spectrum from radio interferometric data. Using the flat
field approximation \citep{datta07}:
\begin{equation}
C_{\ell}=\frac{\sum_{2\pi|\us|=\ell}W(\us)|V(\us)|^2}{\sum_{2\pi|\us|=\ell}W(\us)}
\end{equation}
where $|\us| = \sqrt{u^2 + v^2}$ and $\ell \simeq 2\pi|\us|$ under
flat-field approximation. Here, $V(\u)$ is the un-weighted visibilities
from the residual images and $W(\us)$ is the number of visibilities
entering each $\us$ cell.  

In Figures~\ref{f_ang_src} and~\ref{f_ang_cal}, we have plotted
$\ell(\ell+1)C_{\ell}/(2\pi)$ calculated for one frequency bin of
width 125~KHz from our simulated residual image
maps. Figure~\ref{f_ang_src} shows the angular power spectrum
resulting from using the foreground model that is corrupted by source
position errors.  Figure~\ref{f_ang_cal} illustrates the same result
for the case of residual calibration errors.  The top panels (a) of
both figures show the angular power spectra derived from the UVSUB
residuals of step (iii) in our analysis procedure.  The bottom panels
(b) show the angular power spectra from the final IMLIN residuals
following step (iv).  The vertical lines in panel (b) of both figures
corresponds to $\ell \sim 250$. The IMLIN step, which involves
fitting a third order polynomial over a total bandwidth of 32~MHz, is
expected to remove most of the significant structures for scales
larger than this (corresponding to $\ell\lesssim250$).   All of the
plots have been restricted to $\ell \lesssim 5000$ to match the size
of the MWA synthesized beam (4.5 arcmin).   

The total thermal noise power is much stronger than the angular power
spectra of the \hi 21~cm signal. Hence, we have assumed that the final
power spectrum from the real data will be generated by dividing the
observation into different epochs of equal duration and then
cross-correlating the data cubes from the two epochs
\citep{bowman09}. This approach preserves the persistent \hi 21~cm 
signal and eliminates the thermal noise power (which will be
independent between the two observing epochs and, therefore, average
to zero during the cross-correlation), leaving only the thermal
uncertainty. Hence, the relevant noise figure for the angular power
spectra measurement is given by:

\begin{equation}
C_{\ell}^N = \left <\frac{ \sum_{\ell} \left| N_1(\ell)^{\ast} N_2(\ell) +
N_2(\ell)^{\ast} N_1(\ell) \right|}{2} \right> 
\label{e_anoise}
\end{equation}
where $N_1$ and $N_2$ are simulated noise measurements from two
different epochs \citep{bowman09}. 

The residual angular power spectra in the figures can be compared to
the thermal noise {\it uncertainty} in the observations (Equation~
\ref{e_anoise}) and a fiducial 21~cm signal.  We plot the expected
thermal noise {\it uncertainty} angular power spectrum of the MWA
after 5000 hours of integration assuming a system temperature of
$T_{sys}=250$~K, channel width of $125$~KHz and the observing strategy
described in Section~\ref{s_spec}. The thermal uncertainty spectrum is
shown assuming the angular power spectrum has been binned in
logarithmic intervals of width $\Delta \ell = 0.1$, or approximately
ten bins per decade. For the reference \hi 21~cm signal, we show the
power spectrum for a fully neutral IGM at $z=8$
\citep{furlanetto06}. Modeling the 21~cm signal using a fully neutral
IGM provides a reasonable fiducial expectation since recent 
reionization simulations \citep{lidz08} show that the amplitude of the
power spectrum over the scales probed by the MWA is likely to be even
larger than the fully neutral level when the universe if roughly 50\%
ionized.  It should be noted that different models predict different
amplitudes for the \hi power spectrum. For simplification, we have
used this single realistic model to compare with our residual power
spectrum. The specific conclusions regarding the scale-size dependence
of where residual power will dominate the 21~cm signal will change
depending on the reionization model. 

Figure~\ref{f_ang_src}(a) shows that the angular power spectrum from
the UVSUB images are well above the thermal uncertainty power
spectrum, as well as the model \hi 21~cm signal power spectrum.  In
Figure~\ref{f_ang_src}(b), it is evident that the residual angular
power in the IMLIN image is greatly reduced; and for two of our
fiducial source position error levels ($\sigma_{\theta}=0.1$
and~$0.01$~arcsec), the angular power spectra intercepts the \hi
signal power spectrum at $\ell \lesssim 700$. This shows that the
IMLIN step is very crucial not only for removing faint and confused
continuum foreground sources, but also for removing residual power
left over after subtracting the bright foreground sources.  A source
position accuracy of $\lesssim 0.1$~arcsec would allow the detection
of \hi 21~cm signal at $10 \lesssim \ell \lesssim 2000$ scales. 

Similar features are seen in Figure~\ref{f_ang_cal} for the case of
calibration errors, where the residual angular power spectrum from the
UVSUB image only intercepts the thermal noise power spectrum near
$\ell \sim 2000$ and only the $\sigma_a= 0.01$~$\%$ crosses below the
model \hi 21~cm signal power spectrum and the thermal uncertainty
spectrum. Again, from Figure~\ref{f_ang_cal}(b), it is evident that
the residual angular power spectrum from the IMLIN image is much
lower, particularly below the $\ell=250$ threshold.  We have not
investigated in detail how scales larger than this threshold will be
affected by the polynomial subtraction, but it is likely that some of
the signal will be removed, as well.  A calibration accuracy of
$\sigma_a \lesssim 0.05\%$ should allow the detection of the \hi 21~cm
signal.

\subsection{1D Spherically-Averaged Power Spectrum}
\begin{figure*}[t!]
\centering
\epsfig{file=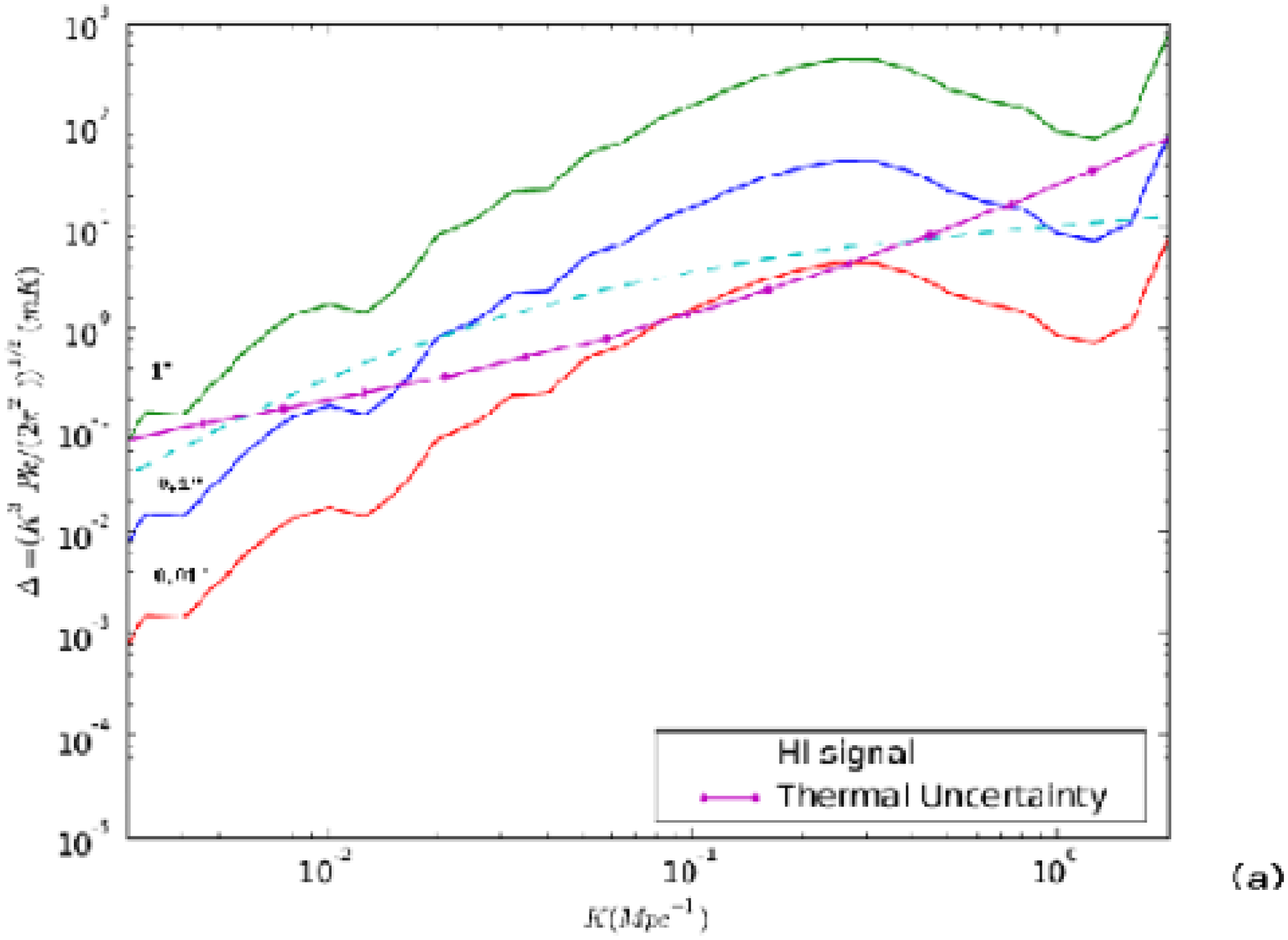,height=3.2truein}
\epsfig{file=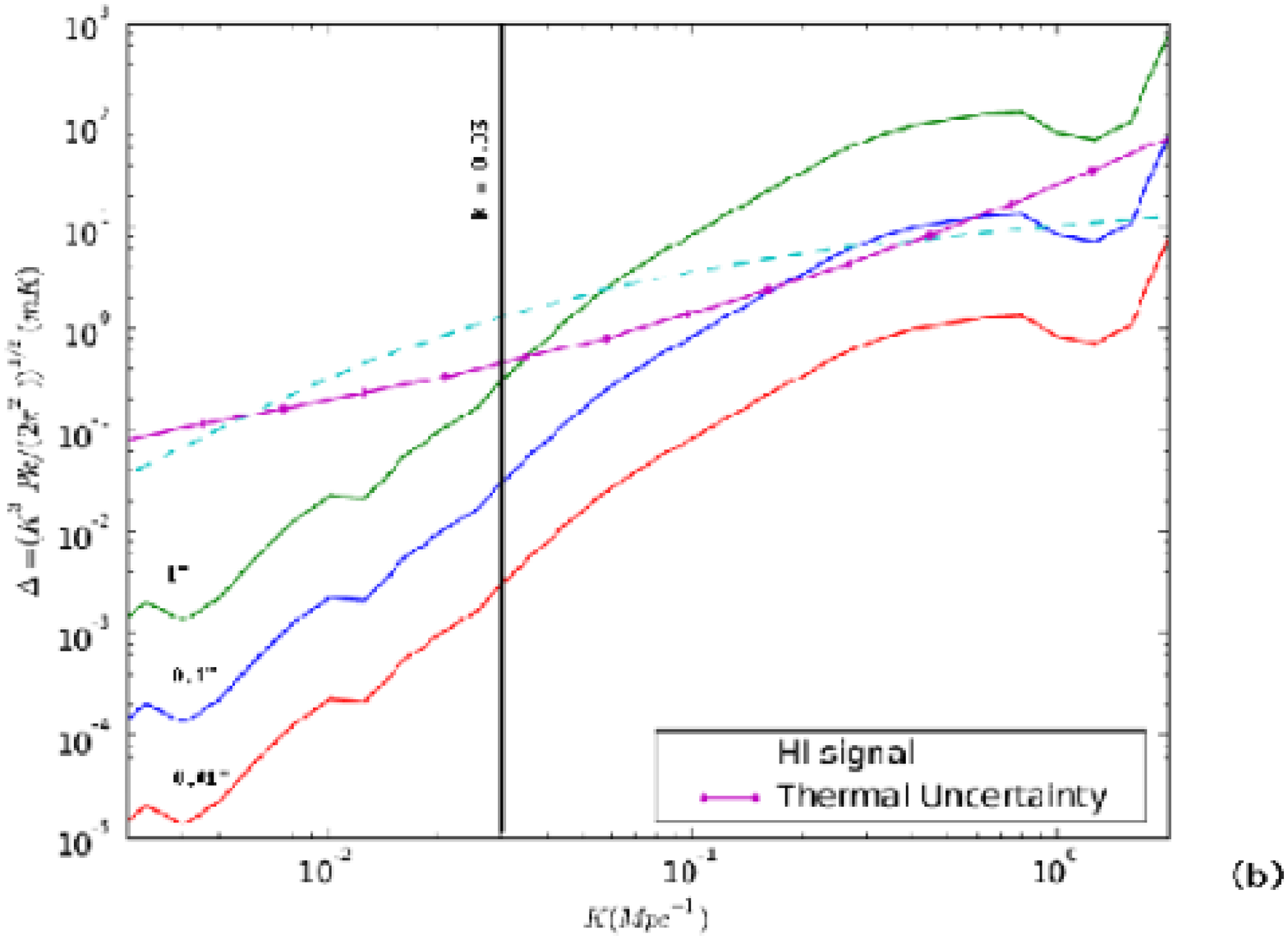,height=3.2truein}
\caption{({\bf a}) 1D spherically-averaged power spectrum of the
  UVSUB residual image $I^{res}(\sc)$ made after subtraction of a
  foreground model with source position errors of
  $\sigma_{\theta}=\{$0.01, 0.1, and 1~arcsec$\}$. ({\bf b}) Same as
  panel (a) but for the IMLIN residual image,
  $I^{res}_{IMLIN}(\sc)$, that is produced after polynomial fitting
  and subtraction has been applied along each sight-line.  The
  vertical line in panel (b) corresponds to $k=0.03$~Mpc$^{-1}$, below
  which the polynomial fitting is expected to remove much of the
  structure.  Both panels include the thermal noise {\it uncertainty}
  spectrum assuming 300~hours of observation with the MWA and binning
  into logarithmic spherical shells of width $\Delta k/k=0.5$, or
  approximately five bins per decade. The \hi 21~cm signal power
  spectrum for a fully neutral IGM ($z\approx8$, $x_{\hi}=1$;
  \citet{furlanetto06}) is also shown.  Detecting the
  spherically-averaged 21~cm power spectrum is the primary goal of the
  MWA.}  
\label{f_pow_src}
\end{figure*}

\begin{figure*}[t!]
\centering
\epsfig{file=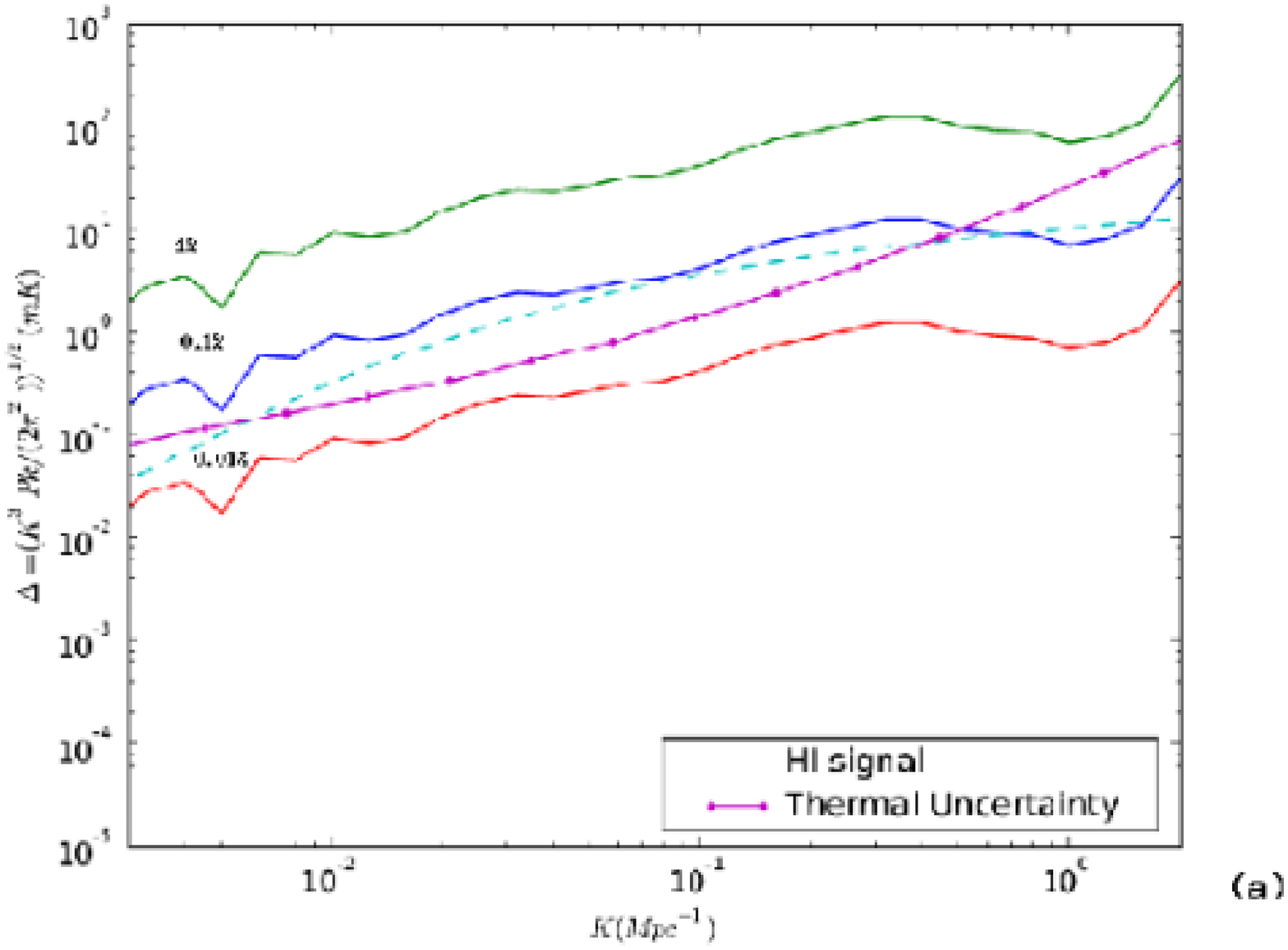,height=3.2truein}
\epsfig{file=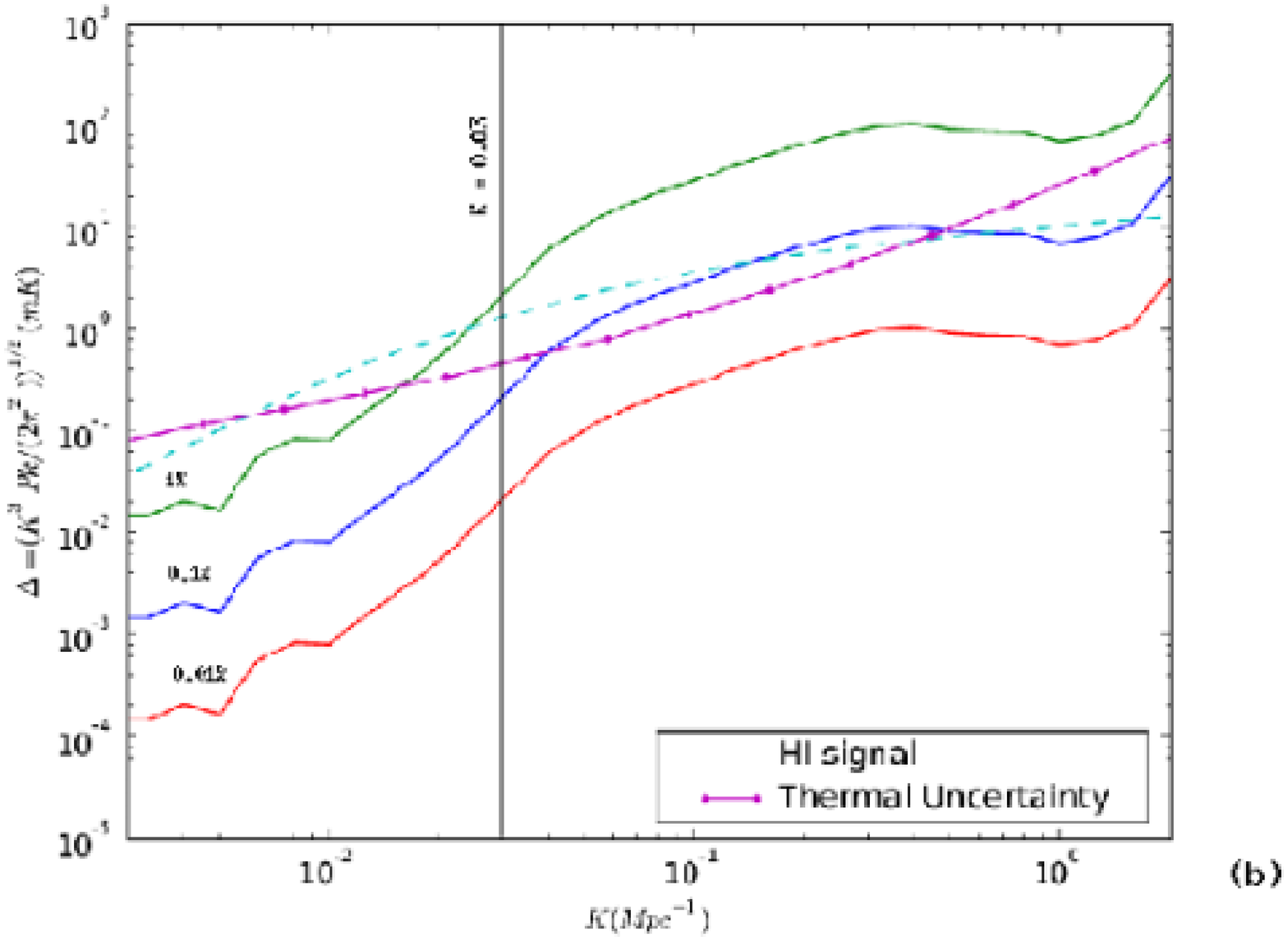,height=3.2truein}
\caption{Same as Figure~\ref{f_pow_src}, but for the residuals due to
  calibration errors.  In this case, the three residual
  spherically-averaged power spectra are for errors of
  $\sigma_a=\{$0.01, 0.1, and 1$\%\}$.}   
\label{f_pow_cal}
\end{figure*}

The spherically-averaged three-dimensional 21~cm power spectrum is the
primary reionization observable targeted by the MWA.  There has been
extensive research on the statistical EoR power spectrum measurement
of the brightness temperature fluctuations in low-frequency,
wide-field radio observations. Detailed formulation has been developed
in the literature by \citet{morales04} and \citet{zfh04}. The
approaches described in these efforts are inspired by the techniques
that have been employed successfully for interferometric measurements
of CMB anisotropies \citep{white99, hobson02, myers03}. The primary
approach is to convert the full three-dimensional measurement cube to
a one dimensional power spectrum. 

The first step is to transform our residual image cubes $I(\sc)$ into
$V(\ut)$ by performing a three dimensional Fourier transform denoted
by the operator $\mathbf{F}(\{\ut\},\{\sc\})$. It should be noted here
that before performing the Fourier transform, we have changed the
units of the residual images from flux unit ($Jy~beam^{-1}$) to
brightness temperature unit (mK). Hence, we get: 
\begin{equation}
V(\ut)=\mathbf{F}(\{\ut\},\{\sc\})I(\sc)
\end{equation}
where $\ut \equiv (u,v,\eta)$. Then, we transform the measurement
coordinates $\u$ into the cosmological coordinates $\k$.  
\begin{eqnarray}
V(\k)&=& \mathbf{J}(\k,\{\ut\})V(\ut) \\
&=& \mathbf{J}(\k,\{\ut\})\mathbf{F}(\{\ut\},\{\sc\})I(\sc)
\end{eqnarray}
where $\mathbf{J}(\k,\{\ut\})$ denotes the Jacobian of the coordinate
transformation from $\ut$ (in units of $\lambda$ and $Hz^{-1}$) to
$\k$ (in units of $cMpc^{-1}$). We have mainly followed the
definition in \citet{peebles93} and the formulation detailed in
\citet{morales04}. Hence, we transformed a residual image cube (in sky
coordinates) to a three dimensional residual visibility cube in the
Fourier conjugate coordinates of co-moving Mpc. 

Assuming isotropy of space and ignoring redshift-space distortions
inherent in converting our observed data cube to cosmological
coordinates, the power spectrum can be taken as approximately
spherically symmetric in cosmological $\k \equiv (k_x,k_y,k_z)$
coordinates. Hence, the power spectrum can be approximated to the
square of the $V(\k)$, averaged over spherical shells: 
\begin{equation} 
P(k) = \left < \left | V(\k) \right |^2 \right >_{|\k|=k}.
\label{e_1d}
\end{equation}
Thus, we obtain the one dimensional total power spectrum
\citep{morales04}, or the more common dimensionless power
spectrum given by $\Delta^2 = k^3 P(k) / (2 \pi^2)$.  

While deriving the 1D power spectrum, we have weighted the individual
measurements $\left | V(\k) \right |^2$ by the per cell visibility
contributions. This scheme is similar to the natural weighting scheme
which is applied to the raw visibilities before imaging, and follows
the form:  
\begin{equation}
P(k) = \frac{\sum_{|\k|=k} W_u(\k)\left | V(\k) \right
|^2 }{\sum_{|\k|=k} W_u(\k)},
\label{e_1d_wt}
\end{equation}
where $W_u(\k)$ denotes the total number of visibilities contributing
per $\k$ cell. Here, we should explicitly mention that the $V(\k)$
used in the above equation are the un-weighted visibilities obtained
from the residual images. 

The total thermal noise power is much stronger than the 1D spherically
averaged power spectra of the \hi 21~cm signal. Hence, similar to the
angular power spectrum case, we compare our results with the thermal
noise uncertainty given by: 

\begin{equation}
P^N(k) = \left <\frac{ \sum_{|\k|=k} \left| N_1(\k)^{\ast} N_2(\k) +
N_2(\k)^{\ast} N_1(\k) \right|}{2} \right> 
\label{e_noise}
\end{equation}
where $N_1$ and $N_2$ are simulated noise measurements from two
different epochs \citep{bowman09}. 

Figures~\ref{f_pow_src} and~\ref{f_pow_cal} show the 1D spherically
averaged power spectrum from the residual images.  As with the angular
power spectrum, these figures also show theoretical \hi 21~cm power
spectrum.  However, here, instead of using a total thermal noise power
spectrum as we did for the angular power spectrum plots, we show the
spherically-averaged thermal noise {\it uncertainty} power spectrum
from 300~hours of observation with the MWA, as mentioned in
Equation~\ref{e_noise}.   The thermal uncertainty spectrum is shown
assuming the spherically-averaged power spectrum has been binned in
logarithmic shells of width $\Delta k/k=0.5$, or approximately five
bins per decade.  As discussed in \citet{lidz08}, the MWA-512 will be
sensitive primarily to scales $0.1 \lesssim k
\lesssim1$~Mpc$^{-1}$. The vertical lines on Figures~\ref{f_pow_src}(b)
and~\ref{f_pow_cal}(b) are at $k=0.03$~Mpc$^{-1}$ and indicate the
scales below which the IMLIN polynomial fitting step removes
significant power.  These lines correspond to the $\ell=250$ threshold
in the angular power spectra plots.  The higher end of the $k$ value for the
residual power spectrum is restricted due to the cell size in the
image domain and frequency resolution of the channels in the residual
image-cube. The maximum value of $\k$ is attained along the $k_z$ axis
only, and hence few or no transverse (angular) modes contribute to
the power spectrum at small scales above $k \gtrsim 0.6$~Mpc$^{-1}$
in Figures~\ref{f_pow_src} and~\ref{f_pow_cal}.  The angular
resolution of the MWA of $\sim4.5$~arcmin (synthesized beam)
corresponds to $k \approx 0.6$~Mpc$^{-1}$.  

Figure~\ref{f_pow_src} shows the 1D spherically-averaged power spectra
from the residual images. These are the residual images after the
foreground subtraction in presence of source position errors.
Figure~\ref{f_pow_src}(a) shows that the 1D power spectra from the
UVSUB image is well below the thermal uncertainty power spectrum and
the \hi signal power spectrum. In Figure~\ref{f_pow_src}(b), it is
evident that the residual 1D spherically-averaged power spectra with
$\sigma_{\theta}=0.01$~and~$0.1$~arcsec from the final IMLIN image 
are below the thermal uncertainty power spectrum and the \hi signal
power spectrum. Hence, a source position accuracy of $\sigma_{\theta}
\lesssim 0.1$~arcsec would allow the detection of \hi 21~cm signal
with the MWA.  

Turning to the case of the calibration errors,
Figure~\ref{f_pow_cal}(a) shows that only the 1D power spectrum from
the UVSUB image for calibration error of  $0.01~\%$ is well below
the thermal uncertainty power spectrum and the \hi signal power
spectrum. In Figure~\ref{f_pow_cal}(b), it is evident that the 1D
spherically-averaged power spectra with $\sigma_{a}=0.01$~and~$0.1\%$
from the IMLIN images are below the thermal uncertainty power
spectrum and the theoretical \hi signal power spectrum. Hence, the
residual calibration accuracy of $\sigma_{a}\lesssim 0.05~\%$ would
allow the detection of \hi 21~cm signal. 

In comparison to the angular power spectrum, we can infer that the 1D
spherically-averaged power spectra has a better tolerance for both the
source position and residual calibration errors. This also reflects
the fact that the angular power spectrum has been produced using a
single channel map of 125 KHz, whereas the 1D spherically-averaged
spectrum is produced with the total bandwidth of 32 MHz.

\begin{figure*}[t!]
\centering
\epsfig{file=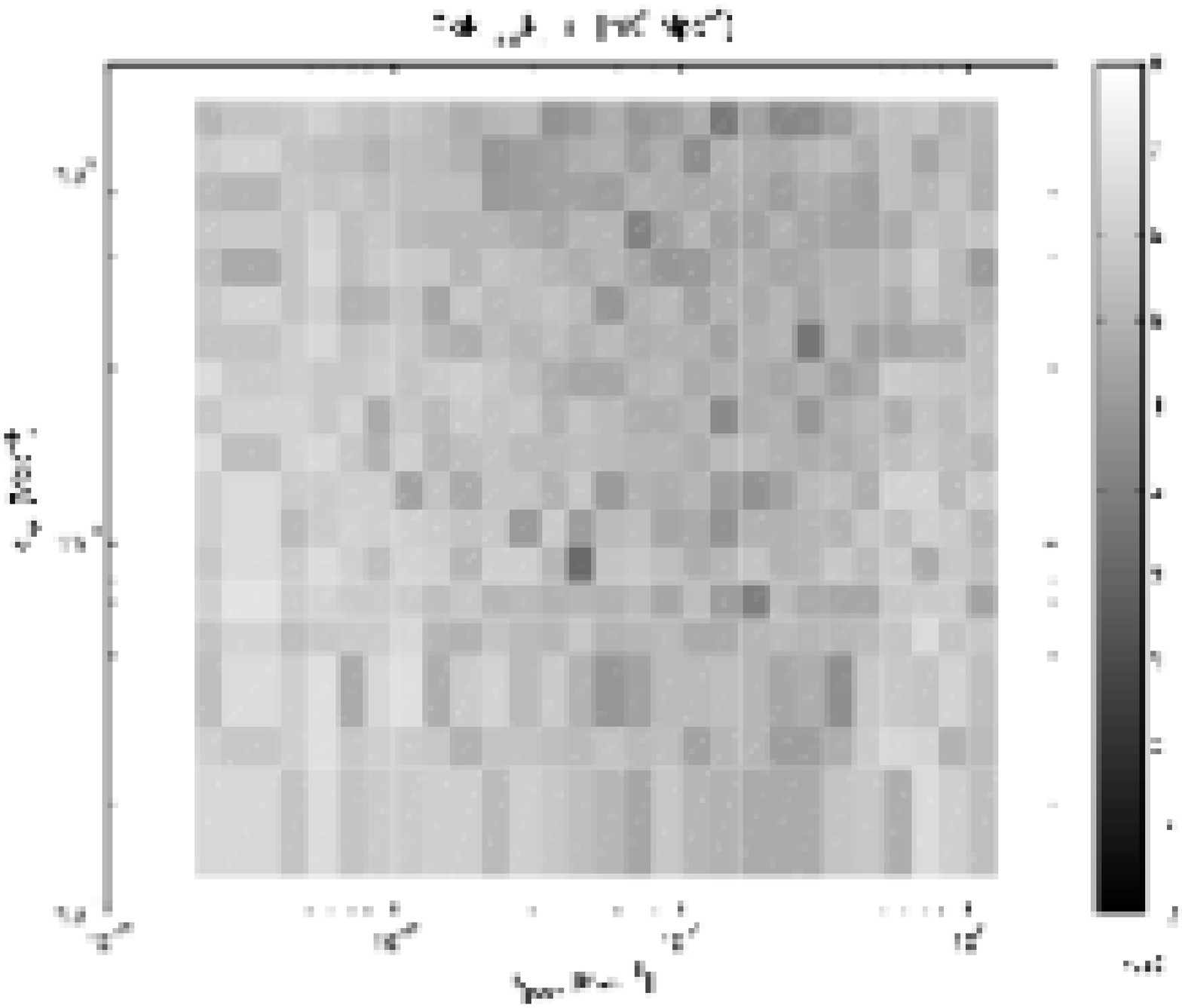,height=2.6truein}
\epsfig{file=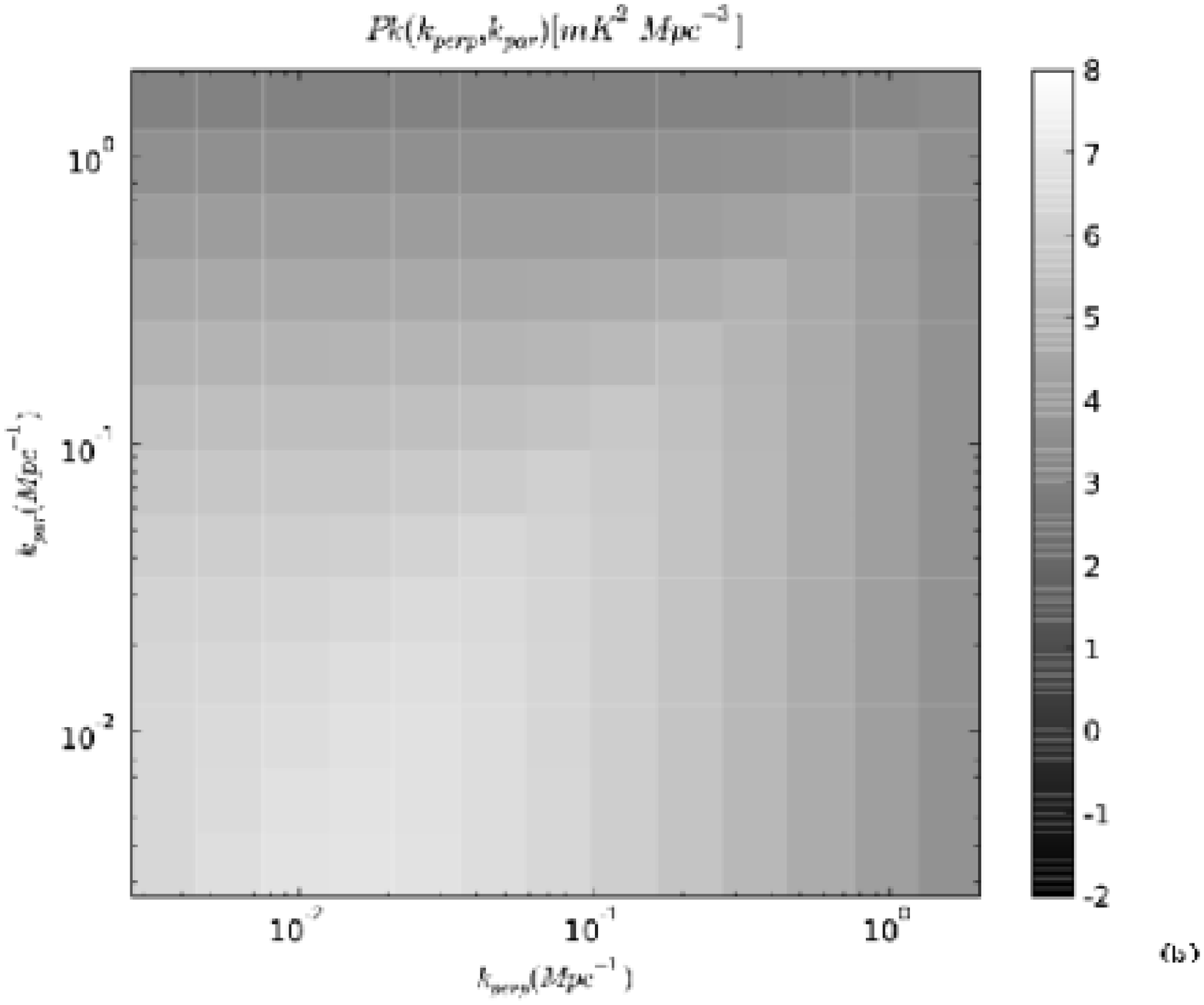,height=2.6truein}
\caption{({\bf a}) Realization of the 2D thermal uncertainty power
  spectrum after cross-correlating simulated thermal noise maps (300
  hours of total integration) from two different epochs
  \citep{bowman09}. ({\bf b}) 2D power spectrum of the \hi 21~cm
  signal \citep{furlanetto06} given by $P(\kperp, \kpar) = (1 + 2 \mu^2
  + \mu^4) P(\k)$, where $\mu = \kpar / |\k|$. Note that the quantity
  plotted here and the following Figures is $P(\kperp, \kpar)$ in
  units of mK$^2$~Mpc$^{-3}$.  The color scale is shown in $\log_{10}
  P(\kperp, \kpar)$.}  
\label{f_2d_ref}
\end{figure*}
\begin{figure*}[t!]
\centering
\epsfig{file=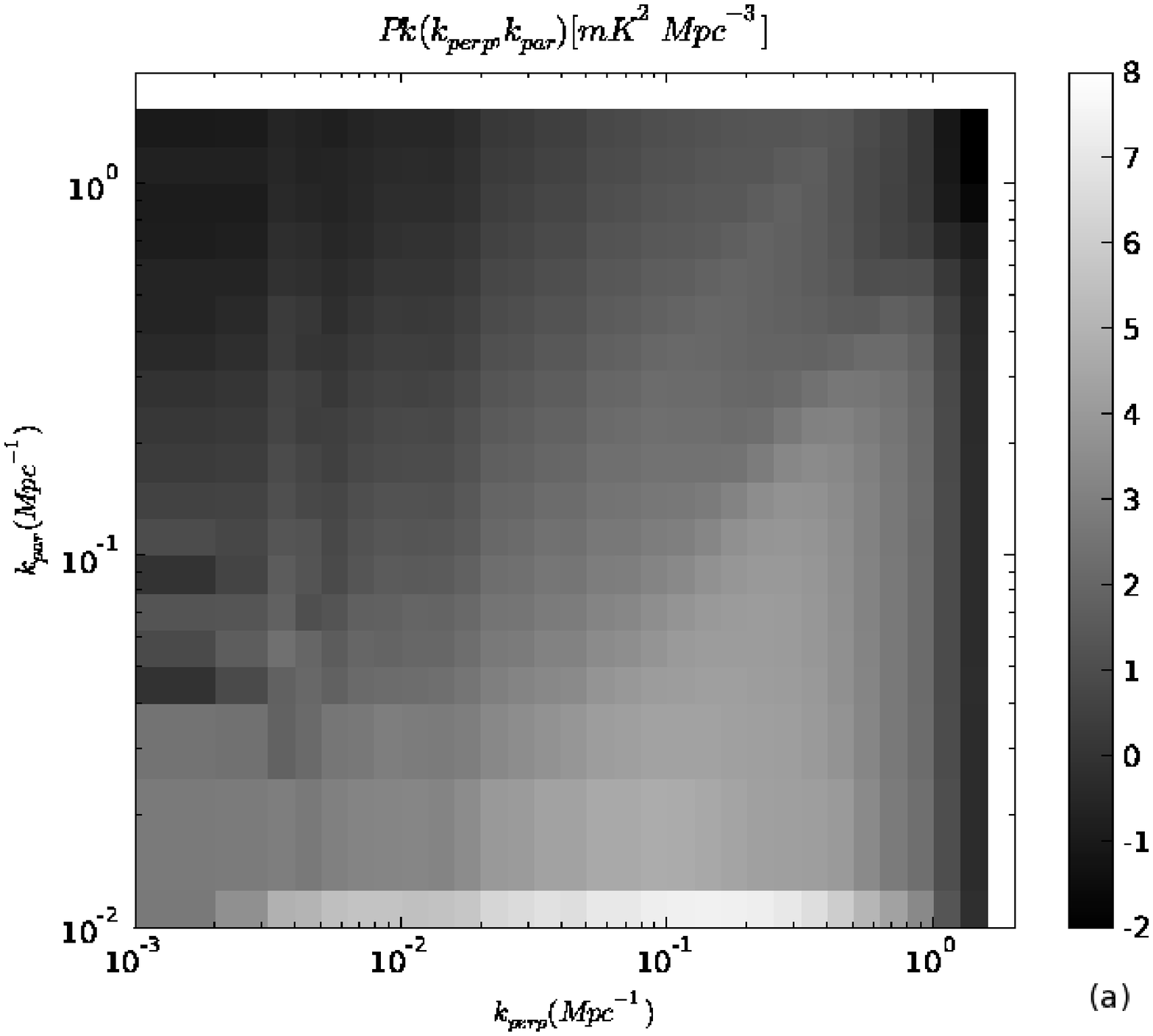,height=2.6truein}
\epsfig{file=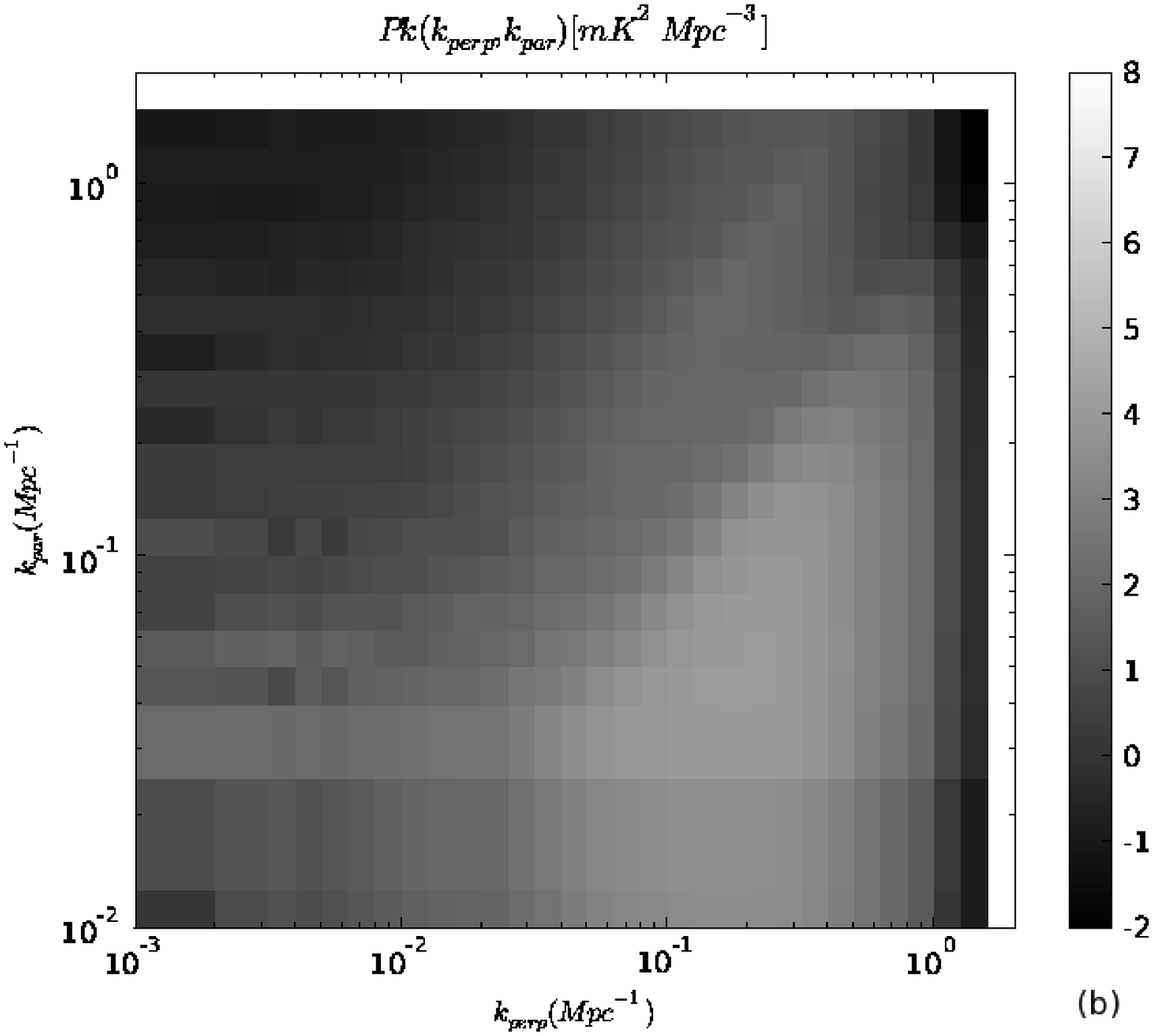,height=2.6truein}
\caption{({\bf a}) 2D power spectrum of the UVSUB residual image
  $I^{res}(\sc)$ made after subtraction of a foreground model with
  source position errors of $\sigma_{\theta}=0.1$~arcsec. ({\bf b})
  Same as panel (a) but for the IMLIN residual image,
  $I^{res}_{IMLIN}(\sc)$, that is produced after polynomial fitting
  and subtraction has been applied along each sight-line. The color
  scale is shown in $\log_{10} P(\kperp, \kpar)$.}  
\label{f_2d_src}
\end{figure*}

\begin{figure*}[t!]
\centering
\epsfig{file=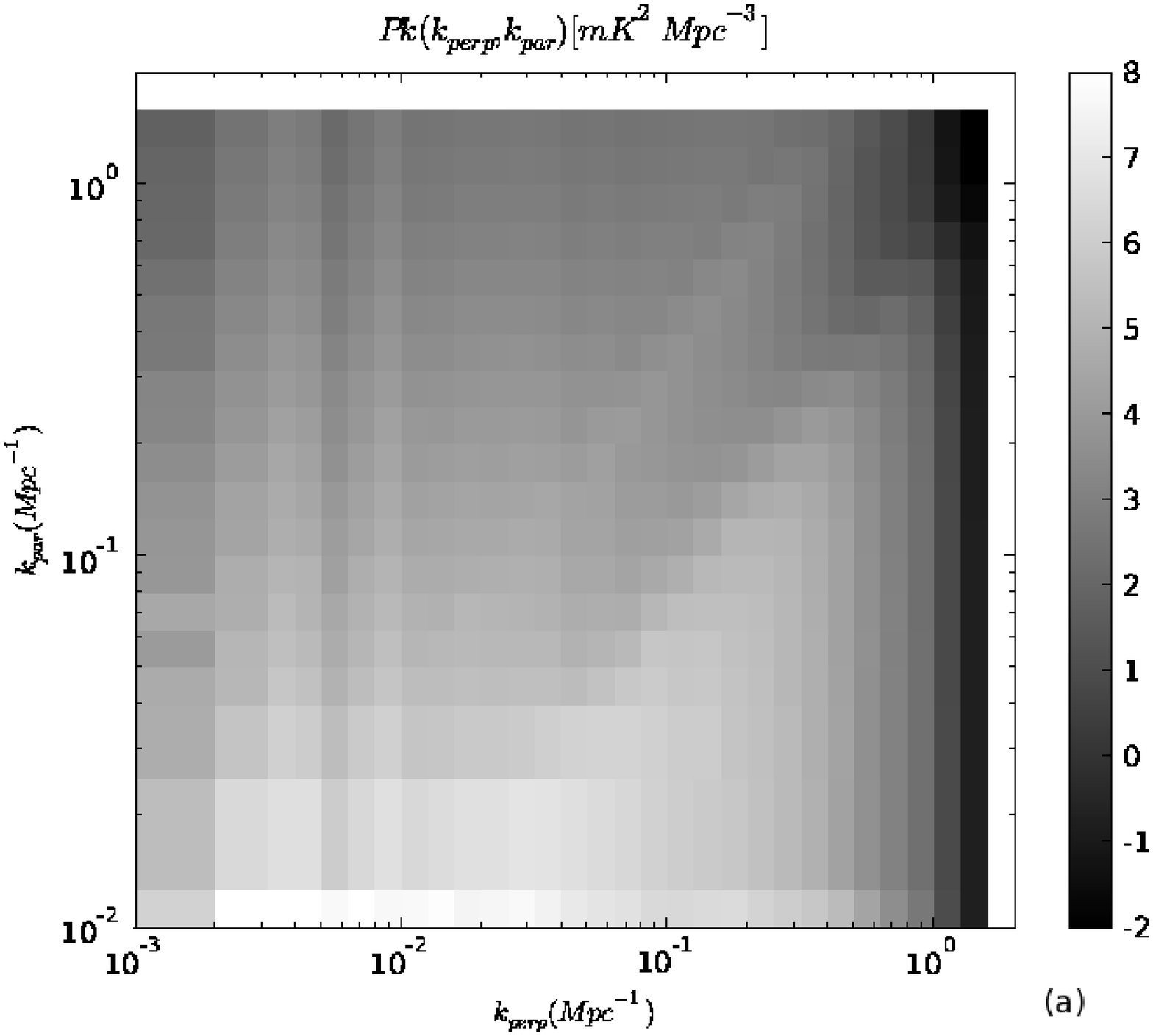,height=2.6truein}
\epsfig{file=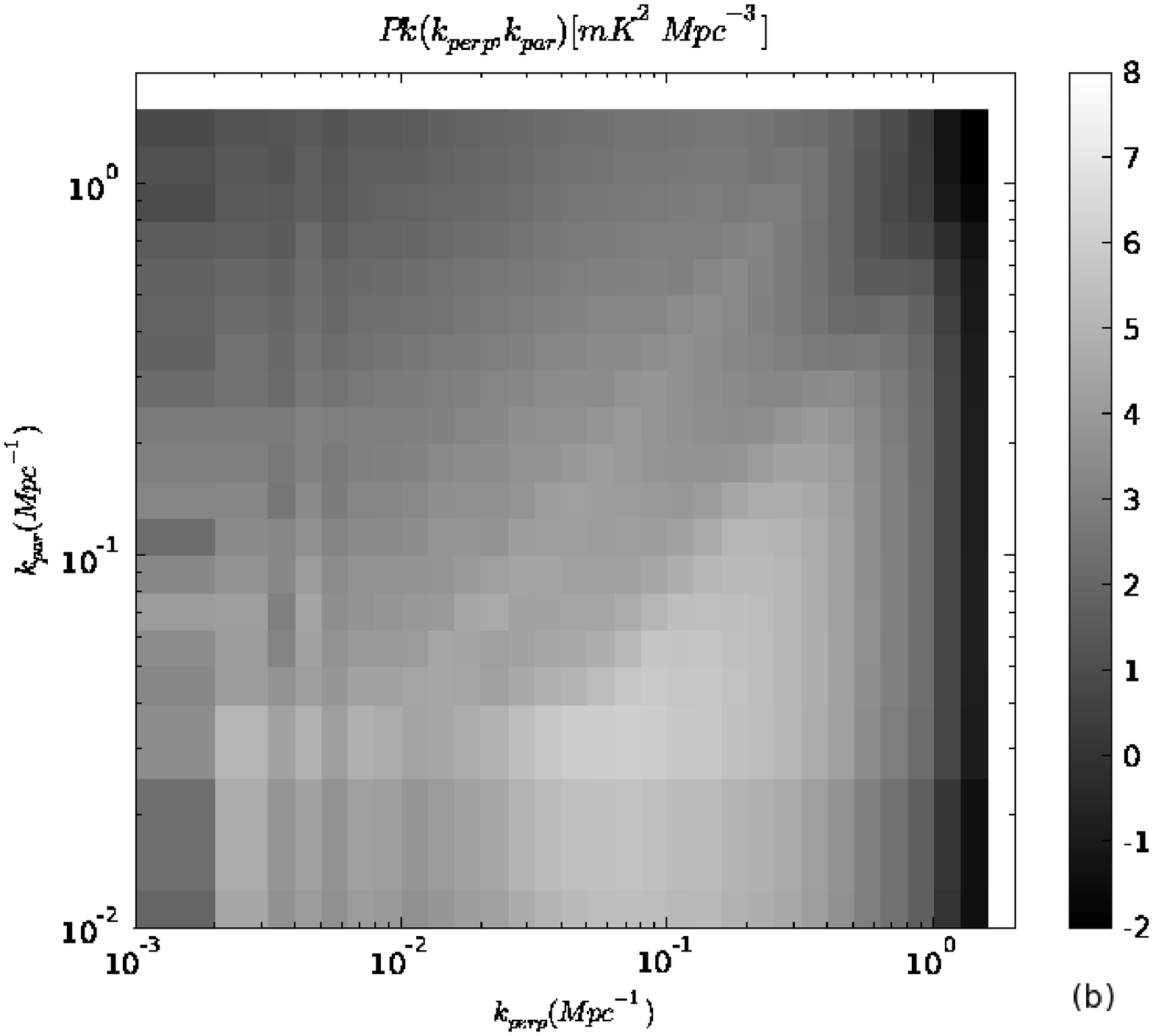,height=2.6truein}
\caption{Same as Figure~\ref{f_2d_src}, but for the residuals due to
  calibration errors.  In this case, the residual 2D power spectrum is
  shown for a fiducial calibration error level of $\sigma_a=0.1\%$. }  
\label{f_2d_cal}
\end{figure*}

\subsection{Two-dimensional Power Spectrum}

In the previous section, we showed the analysis of the 1D
spherically-averaged power spectrum. However, this formulation
mixes the contribution from the $\kperp \equiv \sqrt{k_x^2 +
  k_y^2}$ and $\kpar$ directions. It is useful, therefore, to break
the averaging from the three dimensional $\k$-space to the one
dimensional $k$-space into two steps since both the foregrounds and a
full treatment of the predicted redshifted 21 cm signal that includes
redshifted-space distortions have \textit{aspherical} structure in the
Fourier domain. Following \citet{mcquinn06}, we first average over the
transverse (angular) direction in the full three-dimension power
spectrum to obtain $P(\kperp$, $\kpar)$. This is conceptually similar
to averaging over the $m$ values and keeping the $\ell$ values in a
CMB analysis, except we still have the line-of-sight dimension.  Next,
we obtain the 2D power spectrum based on the maximum likelihood
formalism following the same approach as used for the
spherically-averaged power spectrum in Equations~\ref{e_1d}
and~\ref{e_1d_wt}. 

Figures~\ref{f_2d_ref} through~\ref{f_2d_cal} illustrate the results
of the simulation for the 2D power spectrum.  In all panels of these
figures, the color scale is held constant to facilitate comparison
between the plots.  We show in Figure~\ref{f_2d_ref}(a) a realization
of the 2D thermal noise {\it uncertainty} after 300~hours of
integration with the MWA on our target field. Figure~\ref{f_2d_ref}(b)
shows the 2D \hi signal power spectrum, $P(\kperp, \kpar)$, of the \hi
signal in units of mK$^2$~Mpc$^{-3}$. Figures~\ref{f_2d_src}
and~\ref{f_2d_cal} show the 2D power spectra of the residual image
cubes for source position errors and calibration errors,
respectively. It should be noted that these plots are in different
units than Figure~\ref{f_pow_src} and~\ref{f_pow_cal}.  In this
Section, we have analyzed residual images for only one of our fiducial
error levels for each type of model corruption.  For source position
errors, we use $\sigma_{\theta}=0.1$~arcsec and for residual
calibration errors, we use $\sigma_a=0.1\%$. 

Following our convention, Figure~\ref{f_2d_src}(a) shows that the 2D
power spectrum from the UVSUB image, while Figure~\ref{f_2d_src}(b)
shows the same for the IMLIN image. From Figure~\ref{f_2d_src}(a), we
can conclude that the source position errors are more localized
towards higher $\kperp$ values. Based on the \hi signal shown in
Figure~\ref{f_2d_ref}(b), we see that the signal dominates over 
the residuals at $\kperp \lesssim 0.02$~Mpc$^{-1}$. For the IMLIN
image in Figure~\ref{f_2d_src}(b), the power spectrum shows that \hi
signal dominates at $\kperp \lesssim 0.05$~Mpc$^{-1}$. The thermal
uncertainty map (Figure~\ref{f_2d_ref}(a)) shows that thermal
uncertainty dominates over the source position errors at lower
$\kperp$ values, as mentioned above. There are also regions with high
$\kpar$ (at $\kperp \gtrsim 0.05$~Mpc$^{-1}$) where the thermal
uncertainty dominates over source position errors.  

Figure~\ref{f_2d_cal} shows a very comparable pattern arising from the
residual calibration errors. Hence, we find that the GSM position
accuracy of $0.1$~arcsec and calibration accuracy of $0.1 \%$ is
sufficient to detect the \hi 21~cm signal in 2D power spectrum. The
advantage of the 2D spectrum over the 1D spherically averaged power
spectrum is that one can even search for \hi signal at scales around
$\k \sim 0.1$~Mpc$^{-1}$ along the $\kpar$ axes, which is fairly clean
at lower $\kperp$ values. However, in the 1D power spectrum similar
scales are dominated by the residual errors due to combined
contribution from $\kperp$ and $\kpar$.   
 
\section{Conclusion}

With the results from Section~\ref{s_results}, we can address our
three key questions for bright source subtraction residuals in the
power spectral domain.  First, we find that the level of the source
subtraction accuracy required for power spectral detection of the
21~cm signal is roughly comparable to the accuracy required for direct
imaging of the \hi signal \citep{abhi09}.  The power spectrum
tolerance does suffer, however, compared to the direct imaging case in
one regard.  For direct imaging, it is possible to find areas in the
final image map that are far from bright sources and have very low
residual contamination.  Whereas, for the power spectrum analysis, we
use the entire image map for the calculation, mixing both the good and
the bad areas in the image cube.  Because of this difference, the
power spectrum analysis requires a more stringent calibration accuracy
of $\sigma_a \approx 0.05\%$ compared to $\sigma_a \approx 0.2\%$ for
direct imaging.  This problem is most pronounced for the angular power
spectrum analysis since the residual contamination is dominated by
angular power.  It is mitigated partially in the 1D
spherically-averaged power spectrum by the inclusion of spectral
information--with its lower residual contamination power--in the
calculation, and also in the 2D power spectrum. However, in the case
of the 1D spherically-averaged power spectrum, it should be noted that
the calibration accuracy is determined on the basis of a $\sim
300$~hrs of integration by MWA-512. If we increase the amount of
observing time, the requirement on the accuracy of calibration can be
lowered because our model allows the calibration error to average
toward zero with additional nights of observations. 

The 2-D power spectrum, in particular, addresses our second key
question, showing clear advantages for separating the residual
contamination from the desired signal through distinct localization of
the respective contributions in the the $\kperp$ and $\kpar$
plane. The results for both source position errors and residual
calibration errors indicate that at $\kperp \lesssim 0.05$~Mpc$^{-1}$,
we are able to probe most of the $\kpar$ scales where the \hi signal
is dominant over the residual errors. In the 1D power spectrum we see
dominant contribution from the residual errors around $\k \sim 
0.1$~Mpc$^{-1}$, which can be probed in the 2D power spectra along the
$\kpar$ axes. 

Finally, our third key question was whether we might expect to build a
template library of residual contamination errors in the power
spectrum domain in order to facilitate interpretation of the final
power spectrum.  The results of this work indicate that it will indeed
be possible.  Further, the significant similarities between the 2D
power spectra for the source position error case and the gain
calibration error case suggest that there may be common and easily
identifiable properties of bright source residual contamination that
are largely independent of the specific error causing the
contamination.  This will be a valuable tool for upcoming
experiments. 

We have found that the IMLIN polynomial subtraction step is crucial
not only for faint point source and diffuse foreground subtraction as
studied in other works, but also for the success of the bright source
foreground subtraction that we explored in this paper. 

For the simulations included in this paper, we have performed the
foreground subtraction of bright sources from a data-set of a minimum
of 6~hours of observation (extrapolated to 300 and 5000 hours) in
order to have the full effect of earth rotation synthesis.  However,
the MWA may perform much of its bright source removal over much
shorter timescales ($\sim10$~minutes or less) as part of its real-time
calibration pipeline.  The major implication for shorter time-scale
removal of the foregrounds would be to break our assumption when
modeling source position errors that the position errors are constant
for the entire observation. 

We also made the assumption in this paper that each antenna's
calibration errors are perfectly correlated for an entire 6-hour
observation night, but uncorrelated between observing nights.  If the
residual calibration error where instead perfectly random between {\it
  every} 8-second cycle of the real-time calibration, then we estimate
it could be possible to achieve the desired residual contamination
noise level and detect the redshifted 21~cm $\hi$ signal from
reionization with a significantly larger calibration error of
$\sigma_a~\approx~2.5~\%$. We have not performed our detailed simulation
under this assumption, however, nor have we used the exact parameters
that will be employed for the real-time calibration pipeline of the
MWA. 

The results from the angular power spectrum puts more stringent
contraints on the accuracies in source position and calibration. Using
a larger chunk ($>~ 125 kHz$)of frequency width might have reduced the
constraints. However, we conclude that if a wider bandwidth is
available, it is more advantageous to perform a 1D spherically
averaged power spectrum than the angular power spectrum. This is due
to the inclusion of the $\kpar$ axes contribution to the power
spectrum. 

Lastly, we would like to emphasis that similar constraints can also be
derived for other upcoming arrays, such as LOFAR and PAPER, as well as
for future arrays like the Square Kilometer Array or a lunar
array. But detailed simulations with the unique array specifications
for each instrument would be required, which is beyond the scope of
this paper.  We expect to build on our present analysis in future work
by exploring other arrays, addressing the modified scenarios described
above, and including additional calibration issues such as wide-field
gain calibration of the primary beam and ionosphere.

\acknowledgments 

\noindent AD and CC are grateful for support from the Max-Planck
Society and the Alexander von Humboldt Foundation through the Max
Planck Forshungspreise 2005.  JDB is supported by NASA through Hubble
Fellowship grant HF-01205.01-A awarded by the Space Telescope Science
Institute, which is operated by the Association of Universities for
Research in Astronomy, Inc., for NASA, under contract NAS 5-26555.
The National Radio Astronomy Observatory is operated by Associated
Universities, Inc, under cooperative agreement with the National
Science Foundation.

\bibliographystyle{apj}

\end{document}